\documentclass[11pt,a4paper]{amsart}
\usepackage{amsmath}
\usepackage{amssymb}
\usepackage{amsfonts}
\usepackage[toc,page]{appendix}
\usepackage{hyperref}

\def\={\ =\ }

\newcommand{\be}{\begin{equation}}
\newcommand{\ee}{\end{equation}}
\newcommand{\beq}{\begin{equation}}
\newcommand{\eeq}{\end{equation}}
\newcommand{\bea}{\begin{eqnarray}}
\newcommand{\eea}{\end{eqnarray}}

\def\ba{\begin{eqnarray}}
\def\ea{\end{eqnarray}}

\theoremstyle{plain}

\newtheorem{lemma}{Lemma}

\numberwithin{equation}{section}

\input{tcilatex}

\begin{document}
\title{Supersymmetric $U(N)$ Chern-Simons-matter theory and phase transitions%
}
\author{Jorge G. Russo}
\address{ECM Department and Institute for Sciences of the Cosmos, Facultat
de F\'{\i}sica, Universitat de Barcelona, Mart\'{\i} Franqu\'{e}s 1, E08028
Barcelona, Spain.}
\address{Instituci\'{o} Catalana de Recerca i Estudis Avan\c{c}ats (ICREA),
Pg. Lluis Companys 23, 08010 Barcelona, Spain.}
\email{jorge.russo@icrea.cat}
\author{Guillermo A. Silva}
\address{Departamento de F\'{\i}sica, Universidad Nacional de La Plata \&
Instituto de F\'{\i}sica La Plata, CONICET, C.C. 67, 1900 La Plata,
Argentina.}
\email{silva@fisica.unlp.edu.ar}
\author{Miguel Tierz}
\address{Departamento de An\'{a}lisis Matem\'{a}tico, Universidad
Complutense de Madrid, Plaza de Ciencias 3, 28040 Madrid, Spain.}
\email{tierz@mat.ucm.es}
\maketitle

\begin{abstract}
We study ${\mathcal{N}}=2$ supersymmetric $U(N)$ Chern-Simons with $N_{f}$
fundamental and $N_{f}$ antifundamental chiral multiplets of mass $m$ in the
parameter space spanned by $(g,\,m,\,N,\,N_{f})$, where $g$ denotes the
coupling constant. In particular, we analyze the matrix model description of
its partition function, both at finite $N$ using the method of orthogonal
polynomials together with Mordell integrals and, at large $N$ with fixed $g$%
, using the theory of Toeplitz determinants. We show for the massless case
that there is an explicit realization of the Giveon-Kutasov duality. For
finite $N$, with $N>N_{f}$, three regimes that exactly correspond to the
known three large $N$ phases of theory are identified and characterized.
\end{abstract}



\section{Introduction}


In a classic paper \cite{Mordell}, Mordell analyzed integrals of the type%
\begin{equation}
I=\int_{-\infty }^{\infty }\frac{e^{at^{2}+bt}}{e^{ct}+d}\ dt,  \notag
\end{equation}%
which were originally studied by Kronecker and Lerch in the late 1800s and,
anticipating the comprehensive work by Mordell, they had also appeared in
the study of the Riemann zeta function by Siegel and in relationship with
Mock theta functions by Ramanujan. This latter line of research is of much
current interest after the work \cite{Mock}. This integral also emerges in
studies of unitary representations of extended superconformal algebras (see 
\cite{Eguchi:2010cb} and references therein).

In this paper we consider a quantum field theory where this integral also
plays a central r\^{ }ole and will show that it carries exact
non-perturbative information on the quantum theory. The theory is ${\mathcal{%
N}}=2$ supersymmetric $U(N)$ Chern-Simons (CS) with $N_{f}$ fundamental and $%
N_{f}$ antifundamental chiral multiplets of mass $m$. The partition function
on $\mathbb{S}^{3}$ can be determined by localization techniques \cite%
{Kapustin:2009kz,Kapustin:2010xq,Jafferis:2010un,Hama:2010av} and is given
by 
\begin{equation}
Z_{N_{f}}^{\,U(N)}=\int {d^{N}\!\mu }\frac{\prod_{i<j}4\sinh ^{2}(\frac{1}{2}%
(\mu _{i}-\mu _{j}))\ e^{-\frac{1}{2g}\sum_{i}\mu _{i}^{2}}}{\prod_{i}\left(
4\cosh (\frac{1}{2}(\mu _{i}+m))\cosh (\frac{1}{2}(\mu _{i}-m))\right)
^{N_{f}}}\ ,  \label{Z}
\end{equation}%
where $g=\frac{2\pi i}{k}$ with $k\in \mathbb{Z}$ the Chern-Simons level and 
$\mu _{i}/2\pi $ represent the eigenvalues of the scalar field $\sigma $
belonging to the three dimensional $\mathcal{N}=2$ vector multiplet. In %
\eqref{Z} the radius $R$ of the three-sphere has been set to one. It can be
restored by rescaling $m\rightarrow mR$, $\mu _{i}\rightarrow \mu _{i}R$.
The partition function is periodic in imaginary shifts of the mass, $%
Z(m+i2\pi n)=Z(m)$, for integer $n$.

Localization thus reduces the original functional integral to the
(infinitely) simpler matrix integral (\ref{Z}). However, computing the
remaining $N$ integrations is not straightforward and requires the use of
specific techniques. In the more general case, where the matter chiral
multiplets have $R$-charge $q$ and belong to the representation $R$ of the
gauge group, the matrix model (\ref{Z}) contains double sine functions \cite%
{Jafferis:2010un,Hama:2010av}. A large number of works have been devoted to
analyzing such a matrix model, albeit in a limited region of the parameter
space (e.g. large $N$), whereas in this paper we focus on a more
comprehensive analyzing of (\ref{Z}), which arises when $q=1/2$ and $%
R=r\oplus \overline{r}$. The partition function (\ref{Z}) was calculated in 
\cite{BR} in the large $N$ limit at fixed $gN$ by exactly solving the
saddle-point equations. The planar theory exhibits a number of interesting
features. Non-trivial results emerge when the decompactification limit is
taken by scaling the 't Hooft coupling $t=gN$ with the radius as $%
t=mR\lambda $, with fixed $\lambda $. An inspection of the saddle-point
equations shows that this is the only possible self-consistent scaling that
maintains matter multiplets in the theory (if, instead, $t$ is fixed, then
the decompactification limit just decouples matter multiplets). Thus the
decompactification limit taken in \cite{BR} involves a strong coupling
limit. Then, as $\lambda $ is varied, the theory develops quantum phase
transitions of the third order. The theory presents three phases when $%
0<N_{f}<N$ and two phases for $N_{f}\geq N$. The different phases emerge as $%
\lambda $ is increased from zero: the eigenvalue distribution starts flat
and begins to extend around the origin until it hits $\pm m$. Upon further
increasing of the coupling, the eigenvalues begin to accumulate at $\pm m$
and then, at some higher critical coupling, the boundary of the distribution
overcomes $\pm m$ and continues extending gradually in the form of a flat
distribution with delta function peaks at $\pm m$.

These phase transitions are very similar to phase transitions appearing in
four-dimensional ${\mathcal{N}}=2$ supersymmetric massive gauge theories 
\cite{Russo:2013qaa,Russo:2013kea,Russo:2013sba}. Specifically, this case
parallels phase transitions occurring in four-dimensional ${\mathcal{N}}=2$
Super-QCD \cite{Russo:2013kea}. Recently, similar phase transitions were
found in ${\mathcal{N}}=2$ Chern-Simons theories with bifundamental matter,
such as ABJM and generalizations \cite{Anderson:2014hxa}. Notably, they are
the precise three-dimensional analog of the phase transitions occurring in
the four-dimensional ${\mathcal{N}}=2^{\ast }$ Super Yang-Mills theory \cite%
{Russo:2013qaa,Russo:2013kea}.

The calculation of \cite{BR} showing the existence of phase transitions
describes the CS theory with fundamental matter only in a special corner of
parameter space. Therefore, it is of interest to explore the different
physical and mathematical features of the theory in the complete parameter
space spanned by $(g,\,m,\,N,\,N_{f})$. To this aim, in this paper we will
use different methods to determine the partition function of this theory
first at finite $N$ (by the method of orthogonal polynomials \cite%
{szego,marino}), and then at large $N$ with fixed $g$, by considering the
unitary version of the matrix model, which allows the use of the theory of
Toeplitz determinants \cite{rev,Simon}.

The partition function can also be written, using the change of variables 
\cite{BR} 
\begin{equation}
z_{i}=ce^{\mu _{i}}\ ,\qquad c\equiv e^{g(N-N_{f})}\ ,
\label{changeofvariable}
\end{equation}%
as 
\begin{equation}
Z_{N_{f}}^{\,U(N)}=e^{-\frac{gN}{2}(N^{2}-N_{f}^{2})}\int_{\left[ 0,\infty
\right) ^{N}}{d^{N}z}\ \prod_{i<j}(z_{i}-z_{j})^{2}\frac{e^{-\frac{1}{2g}%
\sum_{i}(\ln z_{i})^{2}}}{\prod_{i}\left( 1+z_{i}\frac{e^{m}}{c}\right)
^{N_{f}}\left( 1+z_{i}\frac{e^{-m}}{c}\right) ^{N_{f}}}\ .  \label{CSM}
\end{equation}%
This ensemble can be formally viewed as a deformation, with logarithmic
potentials, of the Stieltjes-Wigert ensemble whose associated orthogonal
polynomials solve exactly \cite{Tierz} the Chern-Simons matrix model that
describes pure $U(N)$ Chern-Simons theory on $\mathbb{S}^{3}$ \cite%
{Marino:2002fk}. Consideration of the orthogonal polynomial method as
applied to the Hermitian ensemble (\ref{CSM}) leads to the emergence of the
Mordell integral as a crucial tool to obtain explicit analytical expressions
for the partition function. This is developed in Section 2, following an
introduction of the basic formalism of orthogonal polynomials in Section \ref%
{intro}. This use of Mordell integrals not only allows to obtain analytic
expressions for the partition functions, but also provides a very explicit
realization of the Giveon-Kutasov duality \cite%
{Giveon:2008zn,Kapustin:2010mh}, as shown in detail in Section (2.7). In
addition, the existence of such duality, together with the analytical method
developed here, allows to obtain an explicit expression, of the finite-sum
type, for the non-Abelian theory, as shown at the very end of Section 2.

In\ Section 3, we compute $Z_{N_{f}}^{U(N)}$ in a large $g$ limit and with
the mass $m$ also scaling with $g.$ In the large $N$ calculations of \cite%
{BR}, this limit was found to lead to phase transitions and we find here,
for finite $N$, three regimes that are in exact correspondence to the three
large $N$ phases discussed above. Upon taking the large $N$ limit, we will
reproduce the free energies of each phase computed in \cite{BR}.

Then, in Sections 4 and 5, a complementary analysis of the matrix model is
carried out by considering a unitary matrix model version of (\ref{Z}), in
analogy to what occurs in pure Chern-Simons theory \cite{Romo:2011qp}. We
show that the unitary matrix model can be written as%
\begin{equation}
\widetilde{Z}_{N_{f}}^{U(N)}=\left( \frac{g}{2\pi }\right) ^{N/2}\int_{\left[
0,2\pi \right] ^{N}}\frac{d^{N}\mu }{(2\pi )^{N}}\prod_{j=1}^{N}\frac{\theta
_{3}(e^{{i}\mu _{j}},q)}{\left( 4\cos (\frac{1}{2}(\mu _{j}+im))\cos (\frac{1%
}{2}(\mu _{j}-im))\right) ^{N_{f}}}\prod_{i<j}4\sin ^{2}(\frac{1}{2}(\mu
_{i}-\mu _{j})),  \label{unit}
\end{equation}%
where $\theta _{3}(e^{{i}\mu },q)$ is a theta function, and proceed to
analyze it using tools in the theory of Toeplitz determinants \cite%
{rev,Simon}. In particular, we give, using Szeg\"{o}'s theorem, a large $N$
expression for the partition function when $g=2\pi i/k$ is fixed, in
contrast to the large $N$ limit obtained in \cite{BR}, which was taken
keeping $gN$ fixed. In addition, several properties of the matrix model are
presented: (i) the existence of an equivalent matrix model, dual to (\ref%
{unit}) and (ii) the connection between (\ref{unit}) and supersymmetric
versions of Schur polynomials. Both results are generalizations of
properties that also hold for the matrix model that describes pure
Chern-Simons theory \cite{Szabo:2010sd,Szabo:2011eg,Szabo:2013vva}.

In Section 5, we show that more refined results in the theory of Toeplitz
determinants allow one to analyze the massless case \cite{FH,rev}, which is
more delicate to handle than the massive one. In particular, we give an
explicit expression for the partition function for strong-coupling and
finite $N$ and a large $N$ expression for arbitrary coupling.



\section{$U(N)$ partition function from orthogonal polynomials}


\medskip


\subsection{Definitions and conventions}

\label{intro} 

A set of functions $\{\phi_n\}$ satisfying 
\begin{equation}
(\phi_n,\phi_m)=\int \phi_n(x)\phi_m(x) d\alpha(x)=\delta_{nm}
\end{equation}
is said to be orthonormal.

From a set of functions $\{f_{n}\}$ with $n=0,1,2,\ldots $, we can construct
an orthogonal set $\{D_{n}^{(f)}\}$ as follows \cite{szego} 
\begin{equation}
D_{n}^{(f)}(x)=\frac{1}{\mathsf{N_{n}}}\left\vert 
\begin{array}{llll}
(f_{0},f_{0}) & (f_{0},f_{1}) & \cdots & (f_{0},f_{n}) \\ 
(f_{1},f_{0}) & (f_{1},f_{1}) & \cdots & (f_{1},f_{n}) \\ 
~~~\cdots & ~~~\cdots & \cdots & ~~~\cdots \\ 
(f_{n-1},f_{0}) & (f_{n-1},f_{1}) & \cdots & (f_{n-1},f_{n}) \\ 
~~f_{0}(x) & ~~f_{1}(x) & \cdots & ~~f_{n}(x)%
\end{array}%
\right\vert  \label{OP}
\end{equation}%
with 
\begin{equation}
\mathsf{N}_{n}=\left\vert 
\begin{array}{llll}
(f_{0},f_{0}) & (f_{0},f_{1}) & \cdots & (f_{0},f_{n-1}) \\ 
(f_{1},f_{0}) & (f_{1},f_{1}) & \cdots & (f_{1},f_{n}) \\ 
~~~\cdots & ~~~\cdots & \cdots & ~~~\cdots \\ 
(f_{n-1},f_{0}) & (f_{n-1},f_{1}) & \cdots & (f_{n-1},f_{n-1})%
\end{array}%
\right\vert  \label{norm}
\end{equation}%
Here the factor $\mathsf{N}_{n}$ was chosen so that upon choosing ${f}%
_{n}(x)=x^{n}$, the polynomials $p_{n}(x)=D_{n}^{( {f)}}(x)$ have unit
coefficient in its highest power $p_{n}(x)=x^{n}+\ldots $. We define $h_{n}$
as 
\begin{equation}
(p_{n},p_{m})=h_{n}\delta _{nm}\ .  \notag
\end{equation}


\subsubsection*{$U(N)$ Matrix Models}


An Hermitian matrix model has a Jacobian $\Delta
^{2}(z)=\prod_{i<j}(z_{i}-z_{j})^{2}$ arising from gauge fixing the $U(N)$
symmetry 
\begin{equation}
Z=\int d^{N}z\,\Delta ^{2}(z)\,e^{-\frac{1}{g}\sum_{i}V(z_{i})}\ .
\label{PF}
\end{equation}%
The factor $\Delta $, known as Vandermonde determinant, can be written as 
\begin{equation}
\Delta (z)=\left\vert 
\begin{array}{lllll}
1 & z_{1} & (z_{1})^{2} & \cdots  & (z_{1})^{N-1} \\ 
1 & z_{2} & (z_{2})^{2} & \cdots  & (z_{2})^{N-1} \\ 
\cdots  & \cdots  & ~\cdots  & \cdots  & ~\cdots  \\ 
1 & z_{N} & (z_{N})^{2} & \cdots  & (z_{N})^{N-1}%
\end{array}%
\right\vert   \label{vm}
\end{equation}%
Choosing 
\begin{equation}
d\alpha (z)=e^{-\frac{1}{g}V(z)}dz\ ,  \label{measure}
\end{equation}%
as measure, the matrix model orthogonal polynomials $p_{n}(z)$ satisfy 
\begin{equation}
\int p_{n}(z)p_{m}(z)d\alpha (z)=h_{n}\delta _{nm}\ .  \label{pols}
\end{equation}%
A $U(N)$ gauge theory requires the computation of the first $N$ polynomials.
Having the polynomials $p_{n}(z)=z^{n}+\ldots $, and rewriting the
Vandermonde determinant \eqref{vm} as 
\begin{equation}
\Delta (z)=\left\vert 
\begin{array}{lllll}
p_{0}(z_{1}) & p_{1}(z_{1}) & p_{2}(z_{1}) & \cdots  & p_{N-1}(z_{1}) \\ 
p_{0}(z_{2}) & p_{1}(z_{1}) & p_{2}(z_{2}) & \cdots  & p_{N-1}(z_{2}) \\ 
\cdots  & \cdots  & ~\cdots  & \cdots  & ~\cdots  \\ 
p_{0}(z_{N}) & p_{1}(z_{N}) & p_{2}(z_{N}) & \cdots  & p_{N-1}(z_{N})%
\end{array}%
\right\vert =\epsilon ^{i_{1}\ldots i_{N}}p_{i_{1}-1}(z_{1})\cdots
p_{i_{N}-1}(z_{N})\ .  \notag
\end{equation}%
The partition function \eqref{PF} can then be computed as follows \cite%
{marino}: 
\begin{eqnarray}
Z &=&\int d^{N}\alpha (z)\,\Delta ^{2}(z)  \notag \\
&=&\int d^{N}\alpha (z)\,\epsilon ^{i_{1}\ldots i_{N}}\epsilon ^{j_{1}\ldots
j_{N}}p_{i_{1}-1}(z_{1})\cdots p_{i_{N}-1}(z_{N})p_{j_{1}-1}(z_{1})\cdots
p_{j_{N}-1}(z_{N})  \notag \\
&=&\epsilon ^{i_{1}\ldots i_{N}}\epsilon ^{j_{1}\ldots j_{N}}\left( \int
d\alpha (z_{1})\,p_{i_{1}-1}(z_{1})p_{j_{1}-1}(z_{1})\right) \cdots \left(
\int d\alpha (z_{N})\,p_{i_{N}-1}(z_{N})p_{j_{N}-1}(z_{N})\right) \ ,  \notag
\end{eqnarray}%
i.e. 
\begin{equation}
Z=N!\prod_{i=1}^{N}h_{i}\ .  \label{PF2}
\end{equation}%
Alternatively, the partition function $Z$ can be determined from the formula %
\eqref{vm} for $\Delta $. This gives 
\begin{equation}
Z=N!\,\mathsf{N}_{N}\ .  \label{PF33}
\end{equation}

\medskip


\subsection{The ``Mordell" ensemble}


In the present case, we need to compute

\begin{eqnarray}
(\mathsf{f}_{i},\mathsf{f}_{j}) &=&\int d\alpha (z)\,z^{i+j}  \label{fifj} \\
&=&\int_{0}^{\infty }dz\frac{z^{i+j}}{\left( 1+z\frac{e^{+m}}{c}\right)
^{N_{f}}\left( 1+z\frac{e^{-m}}{c}\right) ^{N_{f}}}e^{-\frac{1}{2g}(\ln
z)^{2}}  \notag \\
&=&c^{i+j+1}e^{-\frac{1}{2g}(\ln c)^{2}}\int_{-\infty }^{\infty }d\mu \frac{%
e^{\mu (i+j+1+N_{f}-N)}}{\left( 1+e^{\mu +m}\right) ^{N_{f}}\left( 1+{e^{\mu
-m}}\right) ^{N_{f}}}e^{-\frac{1}{2g}\mu ^{2}}\ .  \notag
\end{eqnarray}%
We have called $z=c\,e^{\mu }$ and $i,j=0,1,\ldots ,N-1$. The second line
reduces to the integral appearing in the Stieltjes-Wigert ensemble by
formally regarding $c$ as an independent parameter and taking $c\rightarrow
\infty $. This is only a formal connection because here $c$ depends on $g$.
In particular, note that 
\begin{equation}
c^{i+j+1}e^{-\frac{1}{2g}(\ln c)^{2}}=e^{g(N-N_{f})\big(i+j+1-\frac{1}{2}%
(N-N_{f})\big)}\ .
\end{equation}%
The partition function is thus given by 
\begin{equation}
Z_{N_{f}}^{\,U(N)}={N!}\ e^{-\frac{gN}{2}(N^{2}-N_{f}^{2})}\ \det (\mathsf{f}%
_{i},\mathsf{f}_{j})\ .  \label{partif}
\end{equation}

\medskip


\subsection{Case $N_f=1$}


\ When $N_f=1$ we can use 
\begin{equation*}
\frac1{(1+a x)(1+b x)}=\frac1{a-b}\left(\frac a{1+a x}-\frac b{1+b x}\right)
\end{equation*}
to obtain 
\begin{equation}
(\mathsf{f}_{i},\mathsf{f}_j)= e^{\frac{g}{2}(N^2-1)}\, e^{\ell g (N-1)}\, 
\frac{I(\ell,m)-I(\ell,-m)}{2\sinh m}\ ,  \label{sprod}
\end{equation}
where the function $I$ is given by 
\begin{equation}  \label{imor}
I(\ell ,m)= \int_{-\infty}^\infty d\mu \frac {e^{(\ell+1) \mu+m}}{ 1+{%
e^{\mu+m}} } e^{-\frac{1}{2g} \mu^2}\ ,
\end{equation}
and the integer $\ell=i+j+1-N$ runs from $\ell =1-N,\ldots,N-1$. Note that
the first exponential factor in (\ref{sprod}) cancels a similar one in %
\eqref{partif} upon taking the determinant.

The integral (\ref{imor}) is a particular case of a Mordell integral \cite%
{Mordell}, which can in general be evaluated in terms of expressions
involving infinite sums. In special cases, the Mordell integrals simplify to
finite Gauss sums, as we shall discuss below and in section \ref{Mordell
Integrals}. For generic values of the parameters, the integral (\ref{imor})
can also be given in terms of infinite sums of error functions, 
\begin{eqnarray}
I(\ell ,m) &=&\int_{-\infty }^{-m}d\mu \frac{e^{(\ell +1)\mu +m}}{1+{e^{\mu
+m}}}e^{-\frac{1}{2g}\mu ^{2}}+\int_{-m}^{\infty }d\mu \frac{e^{(\ell +1)\mu
+m}}{1+{e^{\mu +m}}}e^{-\frac{1}{2g}\mu ^{2}}  \label{I} \\
&=&\sqrt{\frac{\pi g}{2}}\sum_{n=0}^{\infty }(-1)^{n}e^{m(n+1)}e^{\frac{g}{2}%
(n+\ell +1)^{2}}\mathsf{erfc}\left( \frac{g(\ell +n+1)+m}{\sqrt{2g}}\right) 
\notag \\
&&+\sqrt{\frac{\pi g}{2}}\sum_{n=0}^{\infty }(-1)^{n}e^{-mn}e^{\frac{g}{2}%
(n-\ell )^{2}}\mathsf{erfc}\left( \frac{g(n-\ell )-m}{\sqrt{2g}}\right) , 
\notag
\end{eqnarray}%
where $\mathsf{erfc}(x)=1-\mathsf{erf}(x)$ denotes the complementary error
function and we have used $\mathsf{erf}(-x)=-\mathsf{erf}(x)$.

\medskip


\subsection{Case $N_f=1$ and $m=gp$, $p\in\mathbb{N}$}


In this particular case equation \eqref{I} dramatically simplifies and one
obtains 
\begin{equation}
I(\ell,g\,p)=\left\{ 
\begin{array}{ll}
\sqrt{\frac{\pi g}{2}}\,e^{-\frac{gp}{2}(p+2\ell
)}\sum_{n=0}^{2(p+\ell)}(-1)^{n}e^{\frac{g}{2}(p+\ell-n)^{2}}\,, & 
p+\ell\geq 0 \\ 
&  \\ 
\sqrt{\frac{\pi g}{2}}\,e^{-\frac{gp}{2}(p+2\ell
)}\sum_{n=0}^{-2(p+\ell+1)}(-1)^{n}e^{\frac{g}{2}(p+\ell+n+1)^{2}}\,, & 
p+\ell\leq -1%
\end{array}%
\right.  \label{IK}
\end{equation}%
This formula permits the calculation of $Z^{U(N)}$ in terms of elementary
functions, using (\ref{partif}) (in this section, $N_f=1$). In what follows
we give examples for gauge groups $U(1)$, $U(2)$ and $U(3)$.


\subsubsection*{$U(1)$ gauge group:}


\begin{equation}
Z_{p}^{U(1)}=\frac{\sqrt{2 \pi g} \ e^{g p-\frac{g p^2}{2}} } {e^{2 g p}-1}\
\sum _{n=0}^{2 p-1} (-1)^n e^{\frac{1}{2} g (p-n)^2}\ .  \label{ratys}
\end{equation}

\smallskip

\noindent In particular, 
\begin{eqnarray}
&&Z_{p=1}^{U(1)}= \frac{\sqrt{2 \pi g }\ e^{\frac{g}{2}} }{\left(e^{\frac{g}{%
2}}+1\right) \left(e^g+1\right)}  \notag \\
\notag \\
&&Z_{p=2}^{U(1)}=\frac{\sqrt{2 \pi g }\, \left(e^{\frac{3g}{2}}+e^g+e^{\frac{%
g}{2}} -1\right) }{\left(e^{\frac{g}{2}}+1\right) \left(e^g+1\right)
\left(e^{2 g}+1\right)}  \notag \\
\notag \\
&&Z_{p=3}^{U(1)}=\frac{\sqrt{2 \pi g }\ e^{-3 g/2}\ \left(e^{3 g}+e^{\frac{3g%
}{2}}-2 e^{\frac{g}{2}}+1\right) }{\left(e^{\frac{3g}{2}}+1\right)
\left(e^{3 g}+1\right)} \ .  \notag
\end{eqnarray}
Note that the potential pole at $g=0$ in (\ref{ratys}) cancels against a
zero of the numerator.

\medskip

\subsubsection*{$U(2)$ gauge group:}


\begin{eqnarray}
&&Z^{U(2)}_{p=1}= \frac{g\pi\, e^{-g}(e^{\frac{g}{2}}-1)(e^g +2e^{\frac{g}{2}%
}-1)}{(e^{\frac{g}{2}}+1)(e^g+1)}  \notag \\
\notag \\
&&Z^{U(2)}_{p=2}=\frac{g\pi\ e^{-4g} (e^{\frac{g}{2}}-1) (2e^{\frac{3g}{2}%
}+e^g-1) (2e^{5g/2}+e^{2g}-2e^{\frac{g}{2}}+1)}{(e^{\frac{g}{2}}+1)( e^g+1)(
e^{2g}+1)}  \notag \\
\notag \\
&&Z^{U(2)}_{p=3}= \frac{ g\pi e^{-9g} (e^{\frac{g}{2}}-1) (2e^{3g}-e^{g}-e^{%
\frac{g}{2}}+1) (2e^{5g}+2e^{\frac{9g}{2}}-2e^{\frac{7g}{2}%
}+e^{3g}-2e^{2g}+2e^{\frac{g}{2}}-1)} {( e^{\frac{3g}{2}}+1)( e^{3g}+1)} 
\notag
\end{eqnarray}


\subsubsection*{$U(3)$ gauge group:}


\begin{equation}
Z^{U(3)}_{p=1}= \frac{3 \sqrt{2} \, \pi ^{\frac{3}{2}} g^{\frac{3}{2}}\,
e^{-2 g} \left(e^{\frac{g}{2}}-1\right)^3} {e^g+1} \left(e^{\frac{3g}{2}%
}+e^g+e^{\frac{g}{2}}-1\right)  \notag
\end{equation}

\begin{eqnarray}
Z^{U(3)}_{p=2} &=& \frac{3 \sqrt{2} \, \pi ^{\frac{3}{2}} g^{\frac{3}{2}}\,
e^{-6 g} \left(e^{\frac{g}{2}}-1\right)^3 }{\left(e^g+1\right) \left(e^{2
g}+1\right)}\left(e^{2 g}+2 e^{\frac{3g}{2}}-1\right)  \notag \\
&\times & \left(e^{\frac{7g}{2}}+e^{3 g}+e^{\frac{5g}{2}}+e^{2 g} -e^{\frac{%
3g}{2}}-e^g-e^{\frac{g}{2}}+1\right)  \notag
\end{eqnarray}

\begin{eqnarray}
&& Z^{U(3)}_{p=3} =\frac{3 \sqrt{2}\, \pi ^{\frac{3}{2}} g^{\frac{3}{2}}\,
e^{-12 g} \left(e^{\frac{g}{2}}-1\right)^3 \left(2 e^{\frac{7g}{2}}+e^{3
g}-e^{\frac{5g}{2}}-e^{\frac{3g}{2}}-e^{\frac{g}{2}}+1\right) } {%
\left(e^g+1\right) \left(e^g-e^{\frac{g}{2}}+1\right)
\left(e^{2g}-e^g+1\right)}  \notag \\
&&\times \, \left(2 e^{6 g}+e^{\frac{11g}2}+e^{5g}+e^{\frac{9g}2}-e^{4g}-e^{%
\frac{7g}2}-e^{3g}-e^{\frac{5g}2} -e^{2g}+e^{\frac{3g}2}+e^g+e^{\frac{g}%
2}-1\right)  \notag
\end{eqnarray}

It is interesting to interpret these results in terms of the quantized CS
coupling $k$ using $g=2\pi i/k$. In this case the mass, $m=2\pi i p/k $ is
imaginary and the partition function \eqref{Z} depends only on $p$ mod $k$.
From the expressions above we see that for $N=1,2,3$ the partition function $%
Z^{U(N)}_p$ has singularities for particular values of $k$. For example, $%
Z^{U(2)}_{p=3}$ has singularities at $k=1,2,3,6$. In general, the partition
function is regular for $k>2p$. The singularity at $k=2p$ arises because in
this case $m=i\pi $ and the integrand in the partition function \eqref{Z}
acquires a pole on the integration region. The analytic continuation to
imaginary $g$ is therefore only justified for $k>2p$, for $k<2p$ the above
expressions cease to be valid. In the following section we will give general
expressions valid for any integer $k$.


\subsection{Calculation of $Z$ in terms of Mordell integrals}

\label{Mordell Integrals} 

The basic integral $I$ (\ref{imor}) that is used to construct the orthogonal
polynomials has been computed by Mordell \cite{Mordell} for general
parameters. In general, it is given in terms of infinite sums. However, in a
specific case it assumes the form of a Gauss's finite sum. Mordell gives the
remarkable formulas\footnote{%
We correct two small typos on the RHS of (8.2) of \cite{Mordell}
(corresponding to \eqref{GplusM} above): there is no minus sign inside the
square root on the first term, and the limits of summation in the second
term must be shifted by 1.} 
\begin{equation}
\int_{-\infty }^{\infty }dt\ \frac{e^{-i\pi \frac{a}{b}\,t^{2}-2\pi tx}}{%
e^{2\pi t}-1}=G_{-}(a,b,x)\ ,\qquad \int_{-\infty }^{\infty }dt\ \frac{%
e^{i\pi \frac{a}{b}\,t^{2}-2\pi tx}}{e^{2\pi t}-1}=G_{+}(a,b,x)\ ,
\label{wmor}
\end{equation}%
\begin{equation}
G_{-}(a,b,x)\equiv \frac{1}{e^{i\pi b(2x-a)}-1}\left( \sqrt{\frac{-ib}{a}}%
\sum_{r=0}^{a-1}e^{-i\pi \frac{b}{a}(x-r)^{2}}+i\sum_{s=1}^{b}e^{i\pi s(2x+s%
\frac{a}{b})}\right) \ ,  \label{GminusM}
\end{equation}%
\begin{equation}
G_{+}(a,b,x)\equiv \frac{1}{e^{i\pi b(2x-a)}-1}\left( -\sqrt{\frac{ib}{a}}%
\sum_{r=1}^{a}e^{i\pi \frac{b}{a}(x+r)^{2}}+i\sum_{s=0}^{b-1}e^{i\pi s(2x-s%
\frac{a}{b})}\right) \ ,  \label{GplusM}
\end{equation}%
where $a,\,b$ are any positive integers, the square root should be
understood as having positive real part and the integration contour is
deformed to the lower (upper) half plane to avoid the singularity in $G_{-}$(%
$G_{+}$). Notice that $G_{\pm }$ depend only on the ratio $a/b$, even though
it is not manifest in the expressions.

In our case, the integral (\ref{imor}) involves a denominator with positive
relative sign between the two terms. This case can be easily obtained by a
suitable contour deformation as explained in \cite{Mordell}. We find 
\begin{eqnarray}
&&\int_{-\infty }^{\infty }dt\ \frac{e^{-i\pi \frac{a}{b}\,t^{2}-2\pi tx}}{%
e^{2\pi t}-e^{2\pi i\lambda }}=e^{-i\pi (2\lambda +2\lambda x-\frac{a}{b}%
\lambda ^{2})}\ G_{-}\big(a,b,x-\frac{a}{b}\lambda \big)\ ,  \label{qmor} \\
&&\int_{-\infty }^{\infty }dt\ \frac{e^{i\pi \frac{a}{b}\,t^{2}-2\pi tx}}{%
e^{2\pi t}-e^{-2\pi i\lambda }}=e^{i\pi (2\lambda +2\lambda x-\frac{a}{b}%
\lambda ^{2})}\ G_{+}\big(a,b,x-\frac{a}{b}\lambda \big)\ ,  \label{qmor2}
\end{eqnarray}%
where $0\leq \Re (\lambda )<1$. In particular, for $\lambda =1/2$, we get a
denominator with positive relative sign. 
Consider now $I$ (\ref{imor}) with $g=2\pi i/k$, by performing a shift of
integration variable $\mu +m\rightarrow \mu $ and then a rescaling $\mu
\rightarrow 2\pi t$, we can put it into the form 
\begin{equation}
I(\ell ,m)=2\pi \,e^{-m\ell +\frac{ikm^{2}}{4\pi }}\int_{-\infty }^{\infty
}dt\,\frac{e^{i\pi kt^{2}+2\pi t(\ell +1)-itkm}}{e^{2\pi t}+1}\ .
\label{iose}
\end{equation}%
Strikingly, thanks to the fact that $k$ is an integer in CS theory, we can
apply Mordell's formulas \eqref{qmor}-\eqref{qmor2}, which assume that $a$
and $b$ are positive integers. This implies a drastic simplification of the
partition function, since, otherwise, for a generic real number $k$, the
integral $I(\ell ,m)$ would be given by a complicated expression involving
infinite sums.

Thus, using (\ref{qmor2}) with $x=-\ell-1+i k m/2\pi$, $a=k$, $b=1$ and $%
\lambda=1/2$, for any positive integer $k$ we have 
\begin{equation}
I(\ell ,m) = 2\pi \, e^{-i\pi (\ell +\frac{k}{4})}\ e^{-m (\ell+\frac{k}{2})+%
\frac{i k m^2}{4\pi }} G_+ \big(k,1, -\ell-1+i \frac{k m}{2\pi}-\frac{k}{2} %
\big)\ ,  \label{mart}
\end{equation}
with 
\begin{equation}
G_+\big(k,1, -\ell-1+i \frac{k m}{2\pi}-\frac{k}{2} \big) = \frac{1}{e^{ - k
m}-1}\left(- \sqrt{\frac{i }{k}}\ \sum_{r=1}^{k} e^{\frac{i\pi }{k}
(r-\ell-1-\frac{k}{2}+i \frac{k m}{2\pi} )^2}+ i \right)  \label{mert}
\end{equation}
For negative $k$, one must use the analog formula with $G_-$, namely 
\begin{equation}
I(\ell ,m) = 2\pi \, e^{i\pi (\ell -\frac{k}{4})}\ e^{-m (\ell-\frac{k}{2})+%
\frac{i k m^2}{4\pi }} G_- \big(-k,1, -\ell-1+i \frac{k m}{2\pi}+\frac{k}{2} %
\big)\ ,  \label{mart2}
\end{equation}
with 
\begin{equation}
G_-\big(-k,1, -\ell-1+i \frac{k m}{2\pi}+\frac{k}{2} \big) = \frac{1}{%
1-e^{km}}\left(\sqrt{\frac{i}{k}}e^{km} \sum_{r=1}^{-k} e^{\frac{i\pi }{k}
(r+\ell-\frac{k}{2}-i \frac{k m}{2\pi})^2}+ i \right)  \label{mert2}
\end{equation}
We now apply these expressions to compute the $U(N)$ Chern-Simons matter
partition function for arbitrary level and mass. We will make use of the
generalized Gauss's sum identities 
\begin{equation}
\frac{1}{\sqrt{ik}}\sum_{r=1}^{k}e^{\frac{i\pi }{k}(r-\ell -\frac{k}{2}%
)^{2}}=1 \,, \quad\quad~~~~k>0\,,  \label{GI}
\end{equation}%
\begin{equation}
\frac{1}{\sqrt{ik}}\sum_{r=1}^{-k}e^{\frac{i\pi }{k}(r+\ell -\frac{k}{2}%
)^{2}}=1\,, \quad\quad~~~~k<0\,.  \label{GI2}
\end{equation}
valid for $\ell \in \mathbb{Z}$.

It is useful to compare \eqref{mart},\eqref{mart2} with the formulas %
\eqref{IK} for the case $m=gp$. The denominator in (\ref{mert}) becomes
singular for $m=gp=2\pi ip/k$ with integer $p$, however, also the numerator
vanishes in virtue of \eqref{GI}.\footnote{%
One can reverse the logic and use the fact that (\ref{iose}) (and therefore $%
G_{+}$ in (\ref{mert})) is regular at $m=2\pi ip/k$ to actually provide
another proof of the Gauss's identity \eqref{GI}.} By taking the limit $%
p\rightarrow $ integer in \eqref{mart}, and comparing with (\ref{IK}), we
also find the remarkable identity 
\begin{eqnarray}
\quad \quad \quad I(\ell ,gp) &=&\frac{2\pi i}{k}e^{-i\pi (\ell +\frac{k}{4}%
)}\ e^{-\frac{i\pi p}{k}(2\ell +k+p)}\left( p+\frac{k}{2}+\ell +1-\frac{1}{%
\sqrt{ik}}\sum_{r=1}^{k}re^{\frac{i\pi }{k}\left( r-\frac{k}{2}-p-\ell
-1\right) ^{2}}\right)  \label{IK2} \\
&=&\left\{ 
\begin{array}{ll}
\pi \sqrt{\frac{i}{k}}\,e^{-\frac{i\pi p}{k}(p+2\ell )}\sum_{n=0}^{2(p+\ell
)}(-1)^{n}e^{\frac{i\pi }{2k}(p+\ell -n)^{2}}\,, & p+\ell \geq 0 \\ 
&  \\ 
\pi \sqrt{\frac{i}{k}}\,e^{-\frac{i\pi p}{k}(p+2\ell )}\sum_{n=0}^{-2(p+\ell
+1)}(-1)^{n}e^{\frac{i\pi }{k}(p+\ell +n+1)^{2}}\,, & p+\ell \leq -1%
\end{array}%
\right.  \notag
\end{eqnarray}%
valid for $k>0$, and which we verified case by case for various values of $%
p,\ k,\ \ell $.

\smallskip

Summarizing, the partition function can be computed in terms of simple
formulas involving finite sums in two cases: (i) when $m=g p$ for arbitrary
complex number $g$ and integer $p$, or (ii) when $k=2\pi i /g$ is an integer
for arbitrary $m$. The identity \eqref{IK2} ensures that both approaches
agree in the overlapping region of the parameters, that is, when both $%
k=2\pi i /g$ and $p=m/g$ are integers.

In what follows, we give examples for the partition function for arbitrary $%
m $ and different gauge groups. In all cases, $N_{f}=1$.


\subsubsection*{$U(1)$ gauge group:}


In the abelian case, the partition function \eqref{partif} reduces to 
\begin{equation}
Z_k ^{U(1)}=\frac{I(0,m)-I(0,-m)}{2\sinh m}
\end{equation}
from \eqref{mart},\eqref{mart2} we obtain 
\begin{equation}
Z_k ^{U(1)}=\frac{2\pi e^{-m+\frac{i k (m-i \pi )^2}{4 \pi }}}{(1-e^{-2m}) ({%
e^{k m}-1})} \left( \sqrt{\frac i k} \sum _{r=1}^k \left( e^{\frac{i \pi}{k}{%
\ \left(r-1-\frac{k}{2}-\frac{i k m}{2 \pi } \right)^2}} + e^{\frac{i \pi}{k}%
{\left(r-1-\frac{k}{2}+\frac{i k m}{2 \pi} \right)^2}} \right)-2 i \right)
\label{U1k}
\end{equation}
for $k>0$ and 
\begin{equation}
Z_k ^{U(1)}=\frac{2\pi e^{-m+\frac{i k (m+i \pi )^2}{4 \pi }}}{(1-e^{-2m}) ({%
e^{-k m}-1})} \left( \sqrt{\frac i k} \sum _{r=1}^{-k} \left( e^{\frac{i \pi%
}{k}{\ \left(r-\frac{k}{2}-\frac{i k m}{2 \pi } \right)^2+km}} + e^{\frac{i
\pi}{k}{\left(r-\frac{k}{2}+\frac{i k m}{2 \pi} \right)^2-km}} \right)+2 i
\right)  \label{U1k-}
\end{equation}
for $k<0$. These formulas contain perturbative as well as non-perturbative
terms. The perturbative terms arise from the weak-coupling expansion of
factors $e^{\frac{i\pi}{k} (r-1)^2}= e^{\frac{g}{2} (r-1)^2}$, whereas
non-perturbative terms are factors $e^{\frac{i k (m-i \pi )^2}{4 \pi }}=e^{-%
\frac{ (m-i \pi )^2}{2g }}$ and $e^{km}=e^{\frac{ 2\pi i m}{g }}$.

For particular values of $k$, we obtain 
\begin{eqnarray}
&& Z_{(k=1)} ^{U(1)}=\frac{2 \pi e^{m}e^{i\frac\pi4} \left( e^m+1-2 e^{\frac
m2+\frac{im^2}{4\pi}}\right)} {\left(e^m-1\right)^2 \left(e^{m}+1\right)} 
\notag \\
&&Z_{(k=2)} ^{U(1)}= \frac{ \sqrt{2}\pi e^{\frac{i\pi}{4}} e^m \left( e^{2
m}+1 -2 e^{m+\frac{i\pi}{2}} -2\sqrt{2}\, e^{m+\frac{i m^2}{2 \pi }-\frac{%
i\pi}{4}}\right)}{\left(e^{2 m}-1\right)^2}  \notag
\end{eqnarray}


\subsubsection*{$U(2)$ gauge group:}


\begin{eqnarray}
&& Z_{(k=1)}^{U(2)}=\frac{8 i \pi ^2 e^{m+\frac{i m^2}{2 \pi }} \left( e^m+1
-2 e^{\frac{m}{2}-\frac{im^2}{4\pi}}\right)}{\left(e^m-1\right)^2
\left(e^m+1\right)}  \label{mU2k1} \\
&& Z_{(k=2) }^{U(2)}= \frac{8 \pi ^2 e^{2 m} \left( e^{\frac{i m^2}{2\pi }%
}-1\right)\left( e^{\frac{i m^2}{2 \pi }}+i\right)}{\left(e^{2 m}-1\right)^2}
\label{mU2k2}
\end{eqnarray}

\subsubsection*{$U(3)$ gauge group:}

\begin{eqnarray}
&& Z_{(k=1)}^{U(3)}=48 \pi ^3 e^{\frac{i m^2}{2 \pi }+\frac{3 i \pi }{4}}
\label{mU3k1} \\
&& Z_{ (k=2) }^{U(3)} = \frac{24\sqrt{2}\pi ^3 e^{\frac{i\pi}{4}}\, e^{m+%
\frac{ i m^2}{ \pi }}}{\left(e^{2 m}-1\right)^2} \left(e^{2m }+2 i
e^{m}+1-2\sqrt 2 e^{i\frac\pi4} e^{m-\frac{im^2}{2\pi}}\right)  \label{mU3k2}
\end{eqnarray}

\subsection{Massless theory}


The partition function can also be computed in the massless limit. A
convenient way to obtain this case is to consider \eqref{mart},\eqref{mart2}
and take the limit 
\begin{equation}
\lim_{m\rightarrow 0}\frac{I(\ell ,m)-I(\ell ,-m)}{2\sinh m}=\frac{\pi
(-1)^{\ell }}{k^{\frac{3}{2}}}\,e^{\frac{i\pi }{4}(1-k)}\ \sum_{n=0}^{k-1}e^{%
\frac{i\pi }{k}\left( n-\frac{k}{2}-\ell \right) ^{2}}\left( \Big(n-\frac{k}{%
2}\Big)^{2}-\frac{ik}{2\pi }-\ell ^{2}\right) \ ,  \label{sprod2}
\end{equation}%
valid for $k>0$, and 
\begin{equation}
\lim_{m\rightarrow 0}\frac{I(\ell ,m)-I(\ell ,-m)}{2\sinh m}=\frac{\pi
(-1)^{\ell }}{(-k)^{\frac{3}{2}}}\,e^{-\frac{i\pi }{4}(1-k)}\
\sum_{n=1}^{-k}e^{\frac{i\pi }{k}\left( n-\frac{k}{2}+\ell \right)
^{2}}\left( \Big(n+\frac{k}{2}\Big)^{2}-\frac{ik}{2\pi }-\ell ^{2}\right) \ ,
\label{sprod3}
\end{equation}%
for $k<0$.

\noindent Substituting these equations into \eqref{sprod}, \eqref{partif} we
can obtain the partition function in the massless case for any $U(N)$ gauge
group and $N_f=1$.

As an example, we quote the case of $U(1)$ gauge theory: 
\begin{eqnarray}
Z^{U(1)}\bigg|_{m=0} &=&\frac{\pi }{k^{\frac{3}{2}}}e^{\frac{i\pi }{4}%
}\,\sum_{n=0}^{k-1}(-1)^{n}e^{\frac{i\pi }{k}n^{2}}\left( \Big(n-\frac{k}{2}%
\Big)^{2}-\frac{ik}{2\pi }\right)  \notag \\
&=&\frac{1}{2}e^{-\frac{i\pi k}{4}}+\frac{\pi }{k^{\frac{3}{2}}}e^{\frac{%
i\pi }{4}}\,\sum_{n=0}^{k-1}(-1)^{n}e^{\frac{i\pi }{k}n^{2}}\Big(n-\frac{k}{2%
}\Big)^{2},\quad k>0,  \notag
\end{eqnarray}%
\begin{eqnarray}
Z^{U(1)}\bigg|_{m=0} &=& \frac{\pi }{k^{\frac{3}{2}}}e^{\frac{5i\pi }{4}%
}\,\sum_{n=1}^{-k}(-1)^{n}e^{\frac{i\pi }{k}n^{2}}\left( \Big(n+\frac{k}{2}%
\Big)^{2}-\frac{ik}{2\pi }\right)  \notag \\
&=&\frac{1}{2}e^{-\frac{i\pi k}{4}}+\frac{\pi }{k^{\frac{3}{2}}}e^{\frac{%
5i\pi }{4}}\,\sum_{n=1}^{-k}(-1)^{n}e^{\frac{i\pi }{k}n^{2}} \Big(n+\frac{k}{%
2}\Big)^{2} ,\quad k<0,  \label{kneg}
\end{eqnarray}
Generically, the partition function for arbitrary gauge group $U(N)$ in the
massless limit can be obtained from \eqref{U1k}-\eqref{mU3k2} by taking the $%
m\rightarrow 0$ limit.

Finally, note that expressions involving finite sums of the Gauss type are
typical of partition functions in finite quantum mechanics \cite{GQM}. Thus,
it would be interesting to see if the partition function above and also the
massive one \eqref{U1k} can be naturally interpreted as $\mathrm{Tr}\left(
e^{-\beta H}\right) $ over a finite-dimensional Hilbert space.

\subsection{Giveon-Kutasov duality}

In recent years there has been considerable interest in 3d Seiberg-like
dualities \cite{Giveon:2008zn,Kapustin:2010mh}. Our analytical computations
with Mordell integrals allow for an explicit check of such a duality, as we
show in what follows, focussing on the massless case. The duality applies to
the partition function of the type (\ref{Z}) which, written in the same
variables and with the same prefactors as in \cite{Kapustin:2010mh} reads%
\begin{equation}
{\mathcal{Z}}_{N_{f},k}^{U(N)}=\frac{1}{N!}\int {d^{N}\!\lambda }\frac{%
\prod_{i<j}4\sinh ^{2}(\pi (\lambda _{i}-\lambda _{j}))\ e^{\pi
ik\sum_{i}\lambda _{i}^{2}}}{\prod_{i}\left( 4\cosh (\pi (\lambda
_{i}+m))\cosh (\pi (\lambda _{i}-m))\right) ^{N_{f}}}.  \label{duality-Z}
\end{equation}%
In \cite{Kapustin:2010mh} it is shown, in the context of localization and
matrix models, that the Giveon-Kutasov duality of $U(N)$ $\mathcal{N}$=2
Chern-Simons-matter theories \cite{Giveon:2008zn} also holds for $\mathcal{N}
$=3 supersymmetry. More specifically, they find that\footnote{%
In this paper $N_{f}=1$ denotes a pair of fundamental and anti-fundamental
chiral multiplets, therefore there is a factor of 2 relative to the $N_{f}$
of \cite{Kapustin:2010mh}.}%
\begin{equation}
{\mathcal{Z}}_{N_{f},k}^{U(N_{c})}\left( \eta \right) =e^{\mathrm{sgn}(k)\pi
i\left( c_{\left\vert k\right\vert ,N_{f}}-\eta ^{2}\right) }{\mathcal{Z}}%
_{N_{f},-k}^{U(\left\vert k\right\vert +2N_{f}-N_{c})}\left( \eta \right) ,
\label{GK}
\end{equation}%
where the l.h.s. denotes the partition function of a theory with $N_{c}$
colors, $N_{f}$ fundamental chiral multiplets, Chern-Simons level $k$, and a
Fayet-Iliopoulos term $\eta $. The term $c_{\left\vert k\right\vert ,N_{f}}$
is a phase. 
In particular, the matrix model for the case of $N_{f}$ fundamental chiral
multiplets of mass $m$ is considered in \cite{Kapustin:2010mh} and the
duality checked for low values of $N_{f}$.

We will now show that our formulas are consistent with Giveon-Kutasov
duality (\ref{GK}). In particular, for $N_{c}=N_{f}=1$, in the massless case
with $\eta =0$, the duality (\ref{GK}) becomes%
\begin{equation}
{\mathcal{Z}}_{1,-k}^{U(1)}=e^{i\pi \phi (k)}\mathcal{Z}_{1,k}^{U(|k|+1)}\ ,
\label{Mord-duality}
\end{equation}%
where $\phi (k)$ denotes a $k$-dependent phase. One can therefore use this
duality to study large $N_{c}$ limits in terms of a simple integral.

Using \eqref{kneg} we find (recalling that the variables in (\ref{duality-Z}%
) and (\ref{Z}) are related by $2\pi \lambda =\mu $ and a $N!$ prefactor) 
\begin{eqnarray*}
\mathcal{Z}_{1,-1}^{U(1)}& =& \frac{1}{8\pi } e^{\frac{i \pi}{4}} (2-i\pi )
\\
\mathcal{Z}_{1,-2}^{U(1)} &=&\frac{1}{8\pi } e^{\frac{i \pi}{2}} \left( 2
-\left( 1+i\right)\pi \right) \\
\mathcal{Z}_{1,-3}^{U(1)} &=&\frac{1}{72\pi } e^{\frac{i \pi}{4}}\left(
18i+\left( 3-8i\sqrt{3}\right) \pi \right) \\
\mathcal{Z}_{1,-4}^{U(1)} &=&\frac{1}{8\pi } e^{\frac{i \pi}{2}} \left(
2i+(1-2 e^{\frac{i \pi}{4}})\pi \right)
\end{eqnarray*}

We need to compare now with the massless limit of (\ref{mU2k1}) and (\ref%
{mU3k2}), together with a couple of additional higher-rank cases, finding
the highly non-trivial identities 
\begin{eqnarray*}
\lim_{m\to0}\mathcal{Z}_{1,1}^{U(2)} &=& -\frac{1}{8\pi } (2-i\pi ) = e^{%
\frac{3i \pi}{4}} \mathcal{Z}_{1,-1}^{U(1)} \\
\lim_{m\to0}\mathcal{Z}_{1,2}^{U(3)} &=&- \frac{1}{8\pi } \left( 2 -\left(
1+i\right)\pi \right) = e^{\frac{i \pi}{2}} \mathcal{Z}_{1,-2}^{U(1)} \\
\lim_{m\to0}\mathcal{Z}_{1,3}^{U(4)} &=&\frac{1}{72\pi } e^{\frac{4i \pi}{3}%
}\left( 18i+\left( 3-8i\sqrt{3}\right) \pi \right) = e^{-\frac{11 i \pi}{12}%
} \mathcal{Z}_{1,-3}^{U(1)} \\
\lim_{m\to0}\mathcal{Z}_{1,4}^{U(5)}&=& -\frac{1}{8\pi } \left( 2i+(1-2 e^{%
\frac{i \pi}{4}})\pi \right) = e^{\frac{i \pi}{2}} \mathcal{Z}_{1,-4}^{U(1)}
\end{eqnarray*}
Thus, the Giveon-Kutasov dualities are satisfied. From the above relations,
we find the following general expression for the phase: 
\begin{equation}
\phi(k)= \frac{1}{6}+\frac{1}{2}\, k+\frac{7}{12}\, k^2\ .
\end{equation}
Like in \cite{Kapustin:2010mh}, the phase depends quadratically on $k$. The
quadratic ansatz is completely determined by the first three cases $U(2)$, $%
U(3)$, $U(4)$ in the above relations; the last case $U(5)$ is then satisfied
identically. We have checked that the same formula for the phase holds for
higher rank cases.

We stress that the computation of one side of the duality involves the
determinant of an $N_c\times N_c$ matrix of integrals, in particular, a
determinant of a $5\times 5$ matrix in the last line, whereas on the other
side we have a simple integral. More generally, using the duality, we have
derived the formula 
\begin{equation}
\mathcal{Z}_{1,k}^{U(N_c)}\bigg|_{m=0} = e^{-i\pi \phi(k)} \left( \frac{1}{2}%
e^{\frac{i\pi k}{4}}+\frac{\pi }{k^{\frac{3}{2}}}e^{-\frac{i\pi }{4}
}\,\sum_{n=1}^{k}(-1)^{n}e^{-\frac{i\pi }{k}n^{2}} \Big(n-\frac{k}{2}\Big)%
^{2} \right) \ , \qquad k =N_c-1\ .
\end{equation}


\section{Large coupling $g$ limit and phase transitions at large N}


Our starting point is the basic integral (\ref{fifj}) that is used to
compute the determinant $\mathsf{N}_{N}$ 
\begin{equation}
(\mathsf{f}_{i},\mathsf{f}_{j})=g\ e^{g(\ell +N)(N-N_{f})}e^{-\frac{1}{2}%
g(N-N_{f})^{2}}\ J_{ij}\ ,
\end{equation}%
with 
\begin{equation}
J_{ij}=\int_{-\infty }^{\infty }dx\frac{e^{-\frac{g}{2}(x^{2}-2x\ell )}}{%
\left( 4\cosh \frac{1}{2}(gx+m)\cosh \frac{1}{2}(gx-m)\right) ^{N_{f}}}\ ,
\end{equation}%
where $\ell =i+j+1-N$ with $1-N\leq \ell \leq N-1$. The approximations below
rely on the observation that when the coupling $g$ is large, the main
contributions to the integrals come from the saddle-point. 
In this limit the hyperbolic cosine functions in the denominator can be
replaced by exponential functions. We will consider the limit where $m$
scales with $g$, {\textit{i}.e.} $m=gp$, where $p$ is an arbitrary positive
real number. This implies $2\cosh \frac{g}{2}(x\pm p)\rightarrow \exp \frac{g%
}{2}|x\pm p|$. In the large $N$ calculations of \cite{BR}, this limit was
found to lead to phase transitions. We now study the partition function of
the Chern-Simons-matter (CSM) theory in the same limit but for any finite $N$%
, and arbitrary $N_{f}$. It should be noted that this limit is equivalent to
a decompactification limit, since the three-sphere radius appear in the
combination $mR$.

In this limit we thus have 
\begin{eqnarray}
J_{ij}&\approx & \int_{-\infty}^{-p} dx \ e^{-\frac{g}{2} (x^2 -2x(\ell+N_f)
)} +e^{-gp N_f}\int_{-p}^p dx \ e^{-\frac{g}{2} (x^2 -2x\ell )} +
\int_{p}^\infty dx \ e^{-\frac{g}{2} (x^2 -2x(\ell-N_f) )} \ .  \notag
\end{eqnarray}
Which term is dominant depends on the interval where the saddle point lies.
The saddle point at $x=\ell$ in the second term lies inside the interval $%
(-p,p)$ for $p>N-1$ (as $|\ell| \leq N-1$). In this case $J_{ij}$ is just
given by the Gaussian integral of the second term. When $N-N_f-1<p\leq N-1$,
the main contribution comes from the boundaries at $x=\pm p$. Finally, when $%
p\leq N-N_f-1$, the saddle points at $x=\ell\pm N_f$ in the first and third
terms can lie on the intervals $(p,\infty)$ or $(-\infty,-p)$, depending on
the value of $\ell $, in which case the main contributions come from the
first or the third integral.

To keep the discussion general, we may compute analytically the integrals in
terms of error functions. Computing the integrals, we obtain 
\begin{eqnarray}  \label{aret}
J_{ij} &\approx &\sqrt{\frac{g\pi}{2}}\bigg( e^{\frac{g}{2}\ell^2-m N_f} %
\Big( \mathsf{erf}(\frac{m+g\ell}{\sqrt{2g}}) +\mathsf{erf}(\frac{m-g\ell}{%
\sqrt{2g}}) \Big) \\
&+& e^{\frac{g}{2} (\ell+N_f)^2} \ \mathsf{erfc}(\frac{m+g(\ell+N_f)}{\sqrt{%
2g}}) +e^{\frac{g}{2}(\ell-N_f)^2} \ \mathsf{erfc}(\frac{m+g(-\ell+N_f)}{%
\sqrt{2g}}) \bigg) \ .  \notag
\end{eqnarray}
In what follows we will make use of the asymptotic behavior for the error
function $\mathsf{erf}(x) $ at large $|x|$ 
\begin{equation}
\mathsf{erf}(x) \approx \mathrm{sign}(x)- \frac{ e^{-x^2}}{x\sqrt{\pi} }\ ,
\end{equation}
which implies 
\begin{equation}
\mathsf{erfc}(x) =\left\{ 
\begin{array}{ll}
\frac{ e^{-x^2}}{x\sqrt{\pi} } \ ,\qquad & \mathrm{for }\ x>0 \ , \\ 
&  \\ 
2+ \frac{ e^{-x^2}}{x\sqrt{\pi} } \ ,\qquad & \mathrm{for }\ x<0\ .%
\end{array}%
\right.  \notag
\end{equation}
The asymptotic behavior of the error functions in \eqref{aret} depends
crucially on the sign of their arguments. In turn, these depend on $i,j$ and
on the different parameters $g, m, N, N_f$. The strategy is to compute the
determinant by keeping the dominant terms. Notice that for the special case $%
m=g p$ with integer $p$, the argument of the error functions may vanish for
some $i,\ j$ and one has to use $\mathsf{erf}(0)=0 $ instead of the above
asymptotic form.

~

We now discuss the behavior of the partition function as we increase the 't
Hooft coupling $g N$ from 0 to $gN\gg m$. Taking into account that $%
|\ell|\le (N-1)$, we can distinguish three different regimes :

\begin{itemize}
\item[ I.] $0< g< m/({N-1})$: as long as the 't Hooft coupling is bounded by
the mass, the arguments of error functions will always be positive. Then,
the dominant terms are those in the first line of \eqref{aret}, with the sum
of the two error functions replaced by 2. We thus obtain 
\begin{equation}
J_{ij} \approx \sqrt{2\pi g}\ e^{\frac{g}{2}\ell^2-m N_f} \ .  \label{are}
\end{equation}
The partition function, for arbitrary $N,N_f$ and mass $m$ satisfying $%
m>(N-1) g$, to leading order in $g$ results 
\begin{equation}
Z_{N_{f}}^{U(N)} = N!\, e^{-\frac{t}{2} N^2(1-\zeta^2)}\, \mathrm{det }(%
\mathsf{f}_{i},\mathsf{f}_j) \approx N!\, (2\pi g)^{N/2} e^{-m N_{f} N} e^{%
\frac{1}{6}gN(N^2-1)}\ .  \label{zfasei}
\end{equation}
%
Note that the matrix model has become a multiple of the strong coupling
limit of the CS matrix model (see eqn 4.43 in \cite{mputrov})%
\begin{equation}  \label{propCS}
Z_{N_{f}}^{U(N)}=e^{-mN_{f}N}Z_{CS}(\mathbb{S}^{3})\ ,\qquad Z_{CS}(\mathbb{S%
}^{3})\approx N!(2\pi g)^{N/2}e^{\frac{1}{6}gN(N^{2}-1)}\ .
\end{equation}
In the strong coupling limit, the non-trivial Vandermonde term in $Z_{CS}(%
\mathbb{S}^{3})$ simplifies, i.e. $\sinh \left( \left( \mu _{i}-\mu
_{j}\right) /2\right) $ is ``bosonized" to $\exp \left( \left\vert \mu
_{i}-\mu _{j}\right\vert /2\right) $. Therefore, the matrix model for Phase
I is simplified to 
\begin{equation*}
Z=e^{-mN_{f}N}\int d^N\!\mu \,\prod_{i<j}\exp \left( |\mu _{i}-\mu
_{j}|\right) {e }^{-\frac{1}{2g}\sum_i\mu _{i}^{2}}.
\end{equation*}

One can check that the formula (\ref{zfasei}) exactly reproduces the $U(1),\
U(2), \ U(3)$ cases of section 2.4. For $U(3)$ and $U(2)$, the condition $m>
(N-1) g$ is satisfied for $p\ge 3$ and $p\ge2$ respectively; for $U(1)$, it
is always satisfied. The formula (\ref{zfasei}) then arises by keeping the
leading exponentials in the formulas for $U(1),\ U(2), \ U(3) $ of section
2.4.

\medskip

\item[II.] $m/(N-1)\leq g<m/(N-1-N_f) $, with $N_f<N$. In this case, the
arguments of the two error functions in the second line of (\ref{aret}) are
always positive and can be replaced by their asymptotic form $\frac{ e^{-x^2}%
}{x\sqrt{\pi} } $. However, the sign in the argument of the error functions
in the first line of (\ref{aret}) can be positive, negative or zero,
depending on the value of $i+j$. Writing $m=gp$, it can be zero when $N-p-1$
is an even number. As a result, the expression for $Z$ is more involved.

When $N-p-1$ is not an even number, we find 
\begin{equation}
Z_{N_{f}}^{U(N)}=N! \ N_f^{2\beta+2} (2\pi g)^{\frac{N }{2}-1-\beta}\ e^{S}\
\prod_{j=0}^{\beta } \frac{1}{(N -1-p-2j)^2(1+2j-N +N_f+p)^2}\ ,
\end{equation}
with 
\begin{equation}
\beta = \big[\frac{1}{2}(N -p-1)\big] \ ,  \notag
\end{equation}
\begin{eqnarray}
S &=&\frac{1}{6} g \Big(N \left(12 \beta (\beta +2)+6 p \left(2 \beta+2
-N_f\right)+11\right)+N^3-6 (\beta +1) N^2  \notag \\
& & \quad\ -2 (\beta +1) \left(4 \beta (\beta +2)+3 p^2+6 (\beta +1)
p+3\right)\Big)\ ,  \notag
\end{eqnarray}
where ``$[\dots ]$" denotes integer part. Here $p $ is any positive real
number in the interval $N-1-N_f<p\leq N-1$, but with the only condition that 
$N-p-1$ is not even. If $N_f\geq N$, then this regime II extends to
arbitrary low values of $m=g p$.

When $N-p-1$ is even we find 
\begin{equation}
Z_{N_{f}}^{U(N)}=\frac{1}{4}\,N!\ N_{f}^{2\beta }(2\pi g)^{\frac{N }{2}%
-\beta }\ e^{S^{\prime }}\ \prod_{j=0}^{\beta -1}\frac{1}{(N
-1-p-2j)^{2}(1+2j-N +N_{f}+p)^{2}} \ ,  \label{zfaseii}
\end{equation}%
with 
\begin{equation}
m=gp\ ,\qquad \beta =\big[\frac{1}{2}(N -p-1)\big]=\frac{1}{2}(N -p-1)\ , 
\notag
\end{equation}%
\begin{equation}
S^{\prime }=-gpNN_{f}+\frac{1}{6}g\big(3N^{2}+p^{2}-3Np-1\big)\ .  \notag
\end{equation}%
This formula can be compared with the formulas given in the $U(2)$, $U(3)$
case in section 2.4, for $p=1$ and $p=2$ respectively --so that the
condition $g(N-1-N_{f})<m\leq g(N-1)$ is satisfied. Keeping the leading
exponential in $g$, one checks that (\ref{zfaseii}) is exactly reproduced.

\medskip

\item[III.] $m/(N-1-N_{f})\leq g$. This regime exists only when $N_{f}<N$.
Now the arguments of all error functions in (\ref{aret}) may be either
positive or negative according to the value of $i+j$ (or 0, for special
values of $m$ and $i,j$). As a result, $Z$ is complicated also in this case.
For a generic $m=gp$, $0<p\leq N-1-N_{f}$, we obtain 
\begin{equation}
Z_{N_{f}}^{U(N)}=N!\ N_{f}^{2\beta -2\gamma }(2\pi g)^{\frac{N}{2}-\beta
+\gamma }\ e^{I}\ \prod_{j=\gamma +1}^{\beta }\frac{1}{%
(N-1-p-2j)^{2}(1+2j-N+N_{f}+p)^{2}}\ ,  \label{zfaseiii}
\end{equation}%
with 
\begin{equation}
m=gp\ ,\qquad \beta =\big[\frac{1}{2}(N-p-1)\big]\ ,\quad \gamma =\big[\frac{%
1}{2}(N-p-1-N_{f})\big]\ ,
\end{equation}%
\begin{eqnarray}
I &=&\frac{1}{6}g\Big(8\gamma ^{3}+2\gamma \left( -6N\left( N_{f}+p+2\right)
+6(p+2)N_{f}+3N_{f}^{2}+3N^{2}+3p^{2}+12p+11\right)   \notag \\
&+&2\left( 6(p+1)N_{f}+3N_{f}^{2}-\beta \left( 4\beta ^{2}+12\beta
+3p^{2}+6(\beta +2)p+11\right) \right)   \notag \\
&+&N\left( 12\beta ^{2}-6(p+2)N_{f}+12\beta (p+2)-1\right) -12\gamma
^{2}\left( -N_{f}+N-p-2\right) +N^{3}-6\beta N^{2}\Big).  \notag
\end{eqnarray}%
If $N_{f}$ is an even number, then $\gamma =\beta -N_{f}/2$ and the
expression for $I$ simplifies: 
\begin{equation}
I=\frac{1}{6}g\left( -3N^{2}N_{f}-N_{f}\left(
3pN_{f}+N_{f}^{2}+3p^{2}-1\right) +N\left( 3N_{f}^{2}-1\right) +N^{3}\right)
\ ,\qquad N_{f}\ \mathrm{even}.  \notag
\end{equation}%
Similar simplifications can be made for $N_{f}$ odd, leading to formulas
which depend on whether $N$ is even or odd. There are simplifications also
for integer $p$.
\end{itemize}

\medskip

The above three regimes correspond to the three large $N$ phases found in 
\cite{BR}. Adopting the same definition of free energy as in \cite{BR}, 
\begin{equation}
F_{N_{f}}^{U(N)} \equiv -\frac{1}{N^{2}}\ln Z_{N_{f}}^{U(N)}\ ,  \notag
\end{equation}
we can now compare the free energies computed in \cite{BR} for the three
different phases. We take the same Veneziano limit as in \cite{BR}: $%
N\to\infty$, with 
\begin{equation*}
t\equiv gN\ ,\qquad \zeta\equiv \frac{N_f}{N}
\end{equation*}
fixed. 

\bigskip \noindent Phase I) $m> g(N-1)$ case. we now find 
\begin{eqnarray}
F_{N_{f}}^{U(N)} & = &\frac{1}{N^{2}}\left( -\ln N!-\frac{N}{2}\ln (2\pi
g)+N^{2}\zeta m-\frac{1}{6}t(N^{2}-1)\right) \   \notag \\
&\longrightarrow &\ \frac{1}{6}\big(6\zeta m-t\big)\ .  \notag
\end{eqnarray}%
This exactly matches eq. (3.10) of \cite{BR} (in \cite{BR}, $\lambda \equiv
t/m$).

\noindent Phase II) $g(N-1-N_f) <m\leq g(N-1) $. The leading order $O(N^2)$
contribution in $\ln Z$ comes from the exponent $S$. Replacing $\beta $ by $%
(N-p)/2$, and restoring $m$ by $p\to m/g =N/\lambda $, we find 
\begin{equation}
F_{N_{f}}^{U(N)} \approx -\frac{1}{N^2}\ S = \frac{m}{6\lambda^2} \big( %
3(2\zeta-1)\lambda^2+3\lambda -1\big)+O(1/N)\ ,  \notag
\end{equation}
which exactly matches the free energy in the intermediate regime of \cite{BR}%
.

\medskip

\noindent Phase III) $0<m\leq g(N-1-N_f) $. The order $O(N^2)$ contribution
in $\ln Z$ now comes from $I$. Recall $p\to N/\lambda$. At large $N$, we can
replace $\beta\to (N -p)/2$, $\gamma\to (N -p-N_f)/2$. We then find 
\begin{equation}
F_{N_{f}}^{U(N)} = \frac{m}{6\lambda} \big( (\zeta-1)^3\lambda^2+3\zeta^2
\lambda +3\zeta\big)+O(1/N)\ .  \notag
\end{equation}
This exactly matches (3.12) of \cite{BR}.

\medskip

As pointed out in \cite{BR}, the above free energies exhibit discontinuities
in the third derivative with respect to $\lambda $. As in the
four-dimensional case \cite{Russo:2013qaa,Russo:2013kea,Russo:2013sba}, the
discontinuities occur due to resonances produced by extra massless particles
appearing in the spectrum. In the presence of a vev for the scalar field $%
\sigma $ of the vector multiplet, the chiral multiplet masses are
proportional to $|\mu_i\pm m| $. In the large $N$ limit, the matrix integral
is determined by a saddle-point where eigenvalues are distributed
continuously in some interval $(-A,A)$ \cite{BR}. Therefore, extra massless
chiral multiplets contribute to the saddle point when $A$ is greater or
equal $m$. This is the case for $m< gN$, thus producing the discontinuous
behavior in the transition from phase I to phase II. The transition to phase
III --occurring only for $N_f < N$-- seems to be caused by a different
effect: by the time $m$ becomes lower than $g(N-N_f)$, there is a saturation
of $N_f$ eigenvalues located at $\pm m $. In the present context, the origin
of the three regimes can be understood from the changing behavior of matrix
element $J_{ij}$ in the three different intervals, as described above. It is
also worth stressing that for finite $N $ the eigenvalue distribution is not
continuous; the average separation of $\mu_i$ eigenvalues is of $O(1/N)$ and
this is the typical value of a light mass in the spectrum. Thus there are no
sharp resonance effects in this case, unless $N$ is very large.

In conclusion, we have computed the same large $t, m$ limit that in \cite{BR}
led to phase transitions, but now for arbitrary (finite) $N$. Expressions (%
\ref{zfasei}), (\ref{zfaseii}), (\ref{zfaseiii}) for $Z_{N_{f}}^{U(N)}$
apply to any value of $N$ and $N_f$, even low values such as $N=1$ or $N=2$,
they only involve the limit $g\gg 1$, with $m$ scaling with $g$ as $m=gp$
and fixed positive real $p$.


\section{Unitary matrix model formulation and large $N$}


We will now analyze a unitary version of the matrix model, in which the
eigenvalues of the matrix model lie on $\mathbb{S}^{1}$. For pure
Chern-Simons theory on $\mathbb{S}^{3}$ one can indistinctly use the
Hermitian matrix model or the unitary matrix model \cite{Romo:2011qp}. By
considering the unitary version of the matrix model one can employ then
tools from the theory of Toeplitz determinants and also establish
relationships with symmetric functions/polynomials. These relationships
parallel the ones existing for pure Chern-Simons theory, where they are
known to describe some of the connections between Chern-Simons theory and 2d
Yang-Mills theories \cite{Szabo:2013vva}.

Thus, in addition to computing large $N$ free energies for both the massive
and massless cases, we shall establish some mathematical properties
involving supersymmetric versions of Schur polynomials which parallel
results for the pure $U(N)$\ Chern-Simons theory on $\mathbb{S}^{3}$.


\subsection{Toeplitz determinants and Szeg\"{o} theorem}


We begin with a reminder on unitary matrix models through discussion of
their equivalent formulation in terms of Toeplitz determinants and their
computation employing Szeg\"{o}'s theorem. These tools have already been
used in gauge theory in \cite%
{Szabo:2010sd,Szabo:2011eg,Szabo:2013vva,Balasubramanian:2004fz}.

Let $f(z)$ be a complex-valued function on $\mathbb{C}$ with Laurent series
expansion $f(z)=\sum_{k\in \mathbb{Z}}\,f_{k}\,z^{k}$, and let $%
T_{N}(f)=(f_{i-j})_{i,j=1,\dots ,N}$ be the associated Toeplitz operator of
dimension $N$ and symbol $f$. By the Heine-Szeg\"{o} identity, the
corresponding Toeplitz determinant is the partition function of a $U(N)$
unitary matrix model%
\begin{equation}
Z_{N}[f]:=\det T_{N}(f)=\int_{[0,2\pi )^{N}}\frac{d^{N}\phi }{(2\pi )^{N}}%
\prod_{l<k}\,\left\vert e^{\,{i}\phi _{l}}-e^{{i}\phi _{k}}\right\vert ^{2}\
\prod_{j=1}^{N}f(e^{\,{i}\phi _{j}}).  \label{HSid}
\end{equation}%
Notice that the symbol of the Toeplitz determinant is the weight function of
the matrix model and recall that one typically writes $f\left( e^{{\,{i}\,}%
\phi }\right) =\exp \left( -V({e}^{{i}\phi })\right) $ and $V({e}^{{i}\phi
}) $ is the potential of the matrix model. Let $[\ln f]_{k}$, $k\in \mathbb{Z%
}$ denote the coefficients in the Fourier series expansion on the unit
circle $\mathbb{S}^{1}$ of the logarithm of the symbol%
\begin{equation*}
\ln f(z)=\sum_{k=-\infty }^{\infty }\,[\ln f]_{k}\ z^{k}\ ,
\end{equation*}%
and suppose that they obey the absolute summability conditions 
\begin{equation*}
\sum_{k=-\infty }^{\infty }\,\big\vert\left[ \ln f\right] _{k}\big\vert%
<\infty \qquad \mbox{and}\qquad \sum_{k=-\infty }^{\infty }\,k\,\big\vert%
\left[ \ln f\right] _{k}\big\vert^{2}<\infty \ .
\end{equation*}%
Let $\widehat{G}(f)=\exp (\left[ \ln f\right] _{0})$ denote the geometric
mean of the symbol $f$. Then the strong Szeg\"{o} limit theorem for Toeplitz
determinants states~\cite{Simon}%
\begin{equation}
\lim_{N\rightarrow \infty }\,\frac{\det T_{N}(f)}{\widehat{G}(f)}=\exp \Big(%
\,\sum_{k=1}^{\infty }\,k\,\left[ \ln f\right] _{k}\,\left[ \ln f\right]
_{-k}\,\Big)\ .  \label{Sz}
\end{equation}%
Thus the theorem gives an expression for the large $N$ limit of the
partition function (or free energy) of the matrix model in terms of the
Fourier coefficients of the potential. One practical advantage of using
directly this theorem is that one does not need to study the density of
states in the large $N$ (with a saddle-point approximation) in order to
compute the free energy.

We shall focus on the two types of symbol functions that describe pure and
supersymmetric Chern-Simons theory with massive fundamental matter. The
relevant symbol in pure Chern-Simons theory is \cite%
{Szabo:2010sd,Szabo:2013vva}. 
\begin{equation}
\varphi (z)=\prod_{i=1}^{r}(1-x_{i}z)^{-1}(1-y_{i}z^{-1})^{-1}.  \label{S1}
\end{equation}%
The result in \cite{TW} shows that this symbol is dual to the symbol%
\begin{equation}
\widetilde{\varphi }(z)=\prod_{i=1}^{r}(1+x_{i}z)(1+y_{i}z^{-1})  \label{S2}
\end{equation}%
and the corresponding Toeplitz determinants are identical $\det_{N}\left(
\varphi \right) =\det_{N}\left( \widetilde{\varphi }\right) $ (see also \cite%
{Szabo:2010sd,Romo:2011qp}). Notice that the principal specialization%
\footnote{%
In our case, $q=e^{-g}=\exp \left( -2\pi i/k\right) $. That is, there is no
shift $k\rightarrow k+N$ as happens in pure Chern-Simons theory.} $%
x_{i}=y_{i}=q^{i-1/2}$ of the latter directly gives the unitary matrix model
with potential%
\begin{equation}
\exp (-V_{1}(e^{i\theta }))=\lim_{r\rightarrow \infty }\widetilde{\varphi }%
(z;x_{i}=q^{i-1/2},y_{i}=q^{i-1/2})=\frac{\theta _{3}(e^{i\theta },q)}{%
\left( q;q\right) _{\infty }},  \label{o1}
\end{equation}%
with the theta function given by%
\begin{equation}
\theta _{3}({e}^{{i}\theta },q)=\sum_{n=-\infty }^{\infty }q^{n^{2}/2}{e}^{{i%
}n\theta }=\prod\nolimits_{j=1}^{\infty }\left( 1-q^{j}\right) \left( 1+q^{j-%
\frac{1}{2}}{e}^{i\theta }\right) \left( 1+q^{j-\frac{1}{2}}{e}^{-i\theta
}\right) ,  \label{theta}
\end{equation}%
and $\left( q;q\right) _{\infty }=\prod\nolimits_{j=1}^{\infty }\left(
1-q^{j}\right) $ is the $q$-Pochhammer symbol\footnote{%
Obviously, just a nomenclature and not a symbol of a Toeplitz determinant,
like (\ref{S1}) or (\ref{S2}).}. On the other hand, the principal
specialization of the first symbol (\ref{S1}) gives%
\begin{equation}
\exp (-V_{2}(e^{i\theta }))=\lim_{r\rightarrow \infty }\varphi
(z;x_{i}=q^{i-1/2},y_{i}=q^{i-1/2})=\left( q;q\right) _{\infty }\theta
_{3}(-e^{i\theta },q)^{-1}.  \label{o2}
\end{equation}%
Both matrix models have the same partition function, which is essentially
the $U(N)$\ Chern-Simons partition function on $\mathbb{S}^{3}$. It holds
that in the case $r=N$:%
\begin{eqnarray}
Z &=&\int_{(0,2\pi ]^{N}}\frac{d^{N}\mu }{(2\pi )^{N}}\prod_{i<j}4\sin ^{2}(%
\frac{1}{2}(\mu _{i}-\mu _{j}))\ \prod_{j=1}^{N}\varphi (e^{i\mu _{j}})
\label{SC} \\
&=&\int_{(0,2\pi ]^{N}}\frac{d^{N}\mu }{(2\pi )^{N}}\prod_{i<j}4\sin ^{2}(%
\frac{1}{2}(\mu _{i}-\mu _{j}))\ \prod_{j=1}^{N}\widetilde{\varphi }(e^{i\mu
_{j}})  \notag \\
&=&\prod_{i,j}^{N}\frac{1}{1-x_{i}y_{j}}.  \notag
\end{eqnarray}%
The final Cauchy-Binet expression for the two equivalent matrix models in (%
\ref{SC}) follows from Gessel's and Baxter's identities \cite{Bax}, which
are spelled out in detail in the Appendix. From (\ref{SC}) and when $%
x_{i}=y_{i}=q^{i-1/2}$ the matrix models above have a partition function%
\begin{equation}
Z=\prod_{j=1}^{N-1}\frac{1}{\left( 1-q^{j}\right) ^{j}}.  \label{preCS}
\end{equation}%
To obtain the full Chern-Simons partition function the two matrix models
have to be endowed with the right normalization, which is given by the first
factor in the r.h.s. of (\ref{theta}), which is missing in both symbols,
namely the $q$-Pochhammer symbol in its finite version $\left( q;q\right)
_{N}$. Indeed, multiplying the weight function of the two matrix models in (%
\ref{SC}) by $\left( q;q\right) _{N}$ gives a numerical pre-factor $\left(
\left( q;q\right) _{N}\right) ^{N}$ , which manifestly transforms (\ref%
{preCS}) into the Chern-Simons partition function:%
\begin{equation*}
Z_{CS}\left( \mathbb{S}^{3}\right) =\prod_{j=1}^{N-1}\left( 1-q^{j}\right)
^{N-j}.
\end{equation*}%
In principle, Cauchy identity holds when the matrix model is
infinite-dimensional but for this symbol it also holds for the finite case 
\cite{Bax,my}. We show this explicitly in the Appendix, together with the
fact that Szeg\"{o}'s theorem actually corresponds to the Cauchy identity
when the latter is written in terms of Miwa variables.


\subsection{Unitary matrix model and large N}


Let us first write down the trigonometric version for our model
corresponding to supersymmetric CS theory with massive fundamental matter:%
\begin{equation}
\widetilde{Z}_{N_{f}}^{U(N)}=\int_{\left[ -\infty ,\infty \right] ^{N}}\frac{%
d^{N}\mu }{(2\pi )^{N}}\frac{e^{-\frac{1}{2g}\sum_{i}\mu
_{i}^{2}}\prod_{i<j}4\sin ^{2}(\frac{1}{2}(\mu _{i}-\mu _{j}))}{%
\prod_{i}\left( 4\cos (\frac{1}{2}(\mu _{i}+im))\cos (\frac{1}{2}(\mu
_{i}-im))\right) ^{N_{f}}}\ ,  \label{Zmm}
\end{equation}%
making the range of integration compact, as with the pure CS matrix model 
\cite{Romo:2011qp} brings the Gaussian factor into a theta function. Let us
see this explicitly by making the range of integration compact in which case
the weight function is rewritten as follows 
\begin{eqnarray*}
&&\int_{\left[ -\infty ,\infty \right] ^{N}}\prod_{j=1}^{N}\frac{\mathrm{e}%
^{-\frac{1}{2g}\sum_{j=1}^{N}\mu _{j}^{2}}}{\left( 4\cos (\frac{1}{2}(\mu
_{j}+im))\cos (\frac{1}{2}(\mu _{j}-im))\right) ^{N_{f}}}\frac{d\mu _{j}}{%
2\pi }\prod_{i<j}4\sin ^{2}(\frac{1}{2}(\mu _{i}-\mu _{j})) \\
&=&\frac{g^{\frac{N}{2}}}{(2\pi )^{\frac{N}{2}}}\int_{\left[ 0,2\pi \right]
^{N}}\prod_{j=1}^{N}\frac{\sum_{n=-\infty }^{\infty }\mathrm{e}^{-\frac{g}{2}%
n^{2}+in\mu _{j}}}{\left( 4\cos (\frac{1}{2}(\mu _{j}+im))\cos (\frac{1}{2}%
(\mu _{j}-im))\right) ^{N_{f}}}\frac{d\mu _{j}}{2\pi }\prod_{i<j}4\sin ^{2}(%
\frac{1}{2}(\mu _{i}-\mu _{j})) \\
&=&\frac{g^{\frac{N}{2}}}{(2\pi )^{\frac{N}{2}}}\int_{\left[ 0,2\pi \right]
^{N}}\prod_{j=1}^{N}\frac{\theta _{3}({e}^{i\mu _{j}},q)}{\left( 4\cos (%
\frac{1}{2}(\mu _{j}+im))\cos (\frac{1}{2}(\mu _{j}-im))\right) ^{N_{f}}}%
\frac{d\mu _{j}}{2\pi }\prod_{i<j}4\sin ^{2}(\frac{1}{2}(\mu _{i}-\mu _{j})),
\end{eqnarray*}%
where the first equality comes out by expressing the integral over $\left[
-\infty ,\infty \right] $ as an infinite sum of integrals over $\left[
0,2\pi \right] $ while taking into account the periodicity of the
trigonometric functions in the integrand and the identity%
\begin{equation}
\sum_{n=-\infty }^{\infty }\mathrm{e}^{-\beta \left( u+2\pi n\right) ^{2}}=%
\frac{1}{\sqrt{4\pi \beta }}\sum_{n=-\infty }^{\infty }e^{-n^{2}/(4\beta
)}e^{inu},  \label{inv}
\end{equation}%
which follows from Poisson resummation. This allows to make the
identification, in the last equality above, with the theta function (\ref%
{theta}), giving%
\begin{equation}
\widetilde{Z}_{N_{f}}^{U(N)}=\left( \frac{g}{2\pi }\right)
^{N/2}\int_{(0,2\pi ]^{N}}\frac{d^{N}\mu }{(2\pi )^{N}}\frac{\prod_{j}\theta
_{3}(e^{i\mu _{j}},q)\prod_{i<j}4\sin ^{2}(\frac{1}{2}(\mu _{i}-\mu _{j}))}{%
\prod_{i}\left( 4\cos (\frac{1}{2}(\mu _{i}+im))\cos (\frac{1}{2}(\mu
_{i}-im))\right) ^{N_{f}}}\ .  \label{Z2}
\end{equation}%
The denominator can be conveniently factorized as 
\begin{equation}
4\cos (\frac{1}{2}(\mu +im))\cos (\frac{1}{2}(\mu -im))=e^{m}\left(
1+e^{-i\mu }e^{-m}\right) \left( 1+e^{i\mu }e^{-m}\right) .  \label{trig}
\end{equation}%
Hence, we can study the problem from the point of view of Toeplitz
determinants, having to study the symbol:%
\begin{equation}
\varphi _{\mathrm{CSM}}(z)=\frac{\theta _{3}({z},q)}{e^{mN_{f}}\left(
1+e^{-m}/z\right) ^{N_{f}}\left( 1+e^{-m}z\right) ^{N_{f}}}.  \label{symbol}
\end{equation}%
As we shall see below, this type of symbol emerges when studying
supersymmetric Schur polynomials \cite{br}, in the same way the pure CS
matrix model is related to Schur polynomials \cite%
{Szabo:2011eg,Szabo:2013vva}. We will also show below, in (\ref{symbeq}),
that it exists a dual symbol which gives the same partition function.


\subsubsection{Large N limit of the model using Szeg\"{o}'s theorem}


Computation of the Fourier coefficients $[\ln \varphi (z)]_{k}$ and $[\ln
\varphi (z)]_{-k}$ corresponding to (\ref{symbol}) and application of Szeg%
\"{o}'s theorem (\ref{Sz}) gives 
\begin{equation}
\widetilde{Z}_{N_{f}}^{U(N)}=\left( \frac{g}{2\pi }\right) ^{N/2}\frac{%
e^{-NN_{f}\left\vert m\right\vert }}{\left( 1-e^{-2\left\vert m\right\vert
}\right) ^{N_{f}^{2}}}\prod\limits_{j=1}^{\infty }\left( 1-q^{j}\right)
^{N-j}\left( 1-q^{j-\frac{1}{2}}e^{-\left\vert m\right\vert }\right)
^{2N_{f}}\text{ for }N\rightarrow \infty .  \label{massive}
\end{equation}
Note that this is different from the large $N$ limit obtained in \cite{BR},
which was taken keeping $gN$ fixed. Here $g=2\pi i/k$ is fixed.

If we further take the limit of $g\rightarrow \infty $ with $m/g$ fixed,
then (\ref{massive}) reproduces to the expression (\ref{zfasei})
corresponding to phase I. \footnote{%
In the present section the normalization of the partition function differs
by a factor of $(2\pi )^{N}$ from the previous one.} The other phases II and
III cannot be recovered because in the unitary model $|e^{\pm i\mu }e^{-m}|$
is always $<1$ and hence in the large $m$ limit the product of cosine
functions in (\ref{trig}) just reduces to $e^{m}$. As a result, (\ref{Z2})
becomes proportional to the CS matrix partition function model, as in (\ref%
{propCS}).


\subsection{Supersymmetric Schur polynomials}


The mathematical structure involving Schur polynomials and relating
Chern-Simons theory to 2d Yang-Mills theory and its $q$-deformation \cite%
{Aganagic:2004js,Szabo:2013vva}, also appears in our model but with
supersymmetric Schur polynomials \cite{br}. This suggests a relationship
between supersymmetric Chern-Simons theory with massive fundamental matter
and the zero area limit of a supersymmetric version of the combinatorial
Migdal-Witten description of 2d Yang-Mills theory\footnote{%
Two dimensional Yang-Mills theory with a supergroup symmetry, such as $%
U\left( m\right\vert n)$, does not seem to have been previously studied in
the literature. Its extension by substitution of dimensions with
superdimensions might be possible since the supersymmetric Schur polynomials
are known to be characters of both typical and atypical representations \cite%
{MV}. In addition, the related Chern-Simons theory has been extended to the
supergroup setting \cite{Bourdeau:1991hn}.} on $\mathbb{S}^{2}$. This is
similar to the relationship between refined Chern-Simons theory and refined $%
q$-deformed 2d Yang-Mills theory \cite{Gadde:2011uv,Szabo:2013vva} and to
the link found between the superconformal index, which is a twisted
supersymmetric partition function of an $\mathcal{N}$ $=2$ superconformal
field theory on $\mathbb{S}^{3}\times \mathbb{S}^{1}$, and the zero area
limit of $q$-deformed 2d Yang-Mills theory \cite{Gadde:2011ik}. We also
find, as we shall see below, a connection of this type but with a
supersymmetric $q$-deformed version of the dimensions of the zero area 2d
Yang-Mills theory on $\mathbb{S}^{2}$.

We consider the analogue of the expression (\ref{SC}) involving
supersymmetric Schur polynomials $\mathrm{HS}_{\lambda }(x|z)$~\cite{br},
which naturally emerges in the representation theory of Lie superalgebras.
In particular, they are characters of irreducible covariant and
contravariant tensor representations of $\mathfrak{gl}\left( m\right\vert n)$
while Schur polynomials are well-known to be characters in $\mathfrak{gl(}m)$%
. In our setting, we will have $m=N$ and $n=N_{f}$. The polynomials are
defined by \cite{br}%
\begin{equation}
\mathrm{HS}_{\lambda }(x|z)=\sum_{\mu ,\nu }\,N_{\mu \nu }{}^{\lambda
}\,s_{\mu }(x)\,s_{\nu ^{\prime }}(z)  \label{HS1}
\end{equation}%
where $s_{\lambda }(x)$ are Schur polynomials \cite{Mac}, $\lambda ,\,\mu $
and $\nu $ denote representations, indexed by partitions which are
characterized by a sequence of ordered positive numbers, such as $\lambda
=(\lambda _{1},\lambda _{2},...,\lambda _{n})$. The partition $\nu ^{\prime
} $ is the conjugate to $\nu $ and the coefficients $N_{\mu \nu }{}^{\lambda
}\in \mathbb{Z}_{\geq 0}$ are the Littlewood-Richardson coefficients defined
by expressing the ring structure on the space of symmetric polynomials in
the basis of Schur functions as \cite{Mac}%
\begin{equation}
s_{\mu }(x)\,s_{\nu }(x)=\sum_{\lambda }\,N_{\mu \nu }{}^{\lambda
}s_{\lambda }(x)\ ,  \label{LRcoeffs}
\end{equation}%
where the sum is over partitions $\lambda $ of size $\left\vert \mu
\right\vert +\left\vert \nu \right\vert $ \cite{BG}, see Appendix A for
definitions. The Cauchy-Binet identity is now~\cite{br} (see also \cite%
{BG,PG})%
\begin{equation}
\sum_{\lambda }\,\mathrm{HS}_{\lambda }(x|z)\,\mathrm{HS}_{\lambda
}(y|w)=\prod_{i,j\geq 1}\ \frac{(1+x_{i}\,w_{j})\,(1+y_{i}\,z_{j})}{%
(1-x_{i}\,y_{j})\,(1-z_{i}\,w_{j})}\ ,  \label{HSCauch}
\end{equation}%
which we note is symmetric under interchange $(x,y)\leftrightarrow (z,w)$.
We point out that while the sums in (\ref{HS1}) and (\ref{HSCauch}) are
formally over all representations, the size of the partitions that are
summed over is bounded in terms of the number of variables in the symmetric
polynomials, due to the fact that a Schur polynomial is identically $0$ if
the length of its partition is larger than the number of its variables \cite%
{Mac}; see again Appendix A for details. An analogous sum to (\ref{HSCauch})
but with an explicit bound on the size of the first row of $\lambda $ admits
a unitary matrix model description~\cite{PG}%
\begin{eqnarray}
&&\sum_{\lambda ,\lambda _{1}\leq N}\,\mathrm{HS}_{\lambda
}(x_{1},...,x_{k_{1}}|z_{1},...,z_{l_{1}})\,\mathrm{HS}_{\lambda
}(y_{1},...,y_{k_{2}}|w_{1},...,w_{l_{2}})  \label{HS} \\
&=&\int_{[0,2\pi )^{N}}\ \prod_{i=1}^{N}\,\frac{d\phi _{i}}{2\pi }~\,\frac{%
\prod_{j=1}^{k_{1}}\left( 1+x_{j}\,{e}^{i\phi _{i}}\right)
\prod_{j=1}^{k_{2}}\,\left( 1+y_{j}\,e^{-i\phi _{i}}\right) }{%
\prod_{j=1}^{l_{1}}\left( 1-z_{j}\,{e}^{i\phi _{i}}\right)
\,\prod_{j=1}^{l_{2}}\left( 1-w_{j}\,{e}^{-i\phi _{i}}\right) }%
\prod_{i<j}4\sin ^{2}(\frac{1}{2}(\phi _{i}-\phi _{j}))\ .  \notag
\end{eqnarray}%
This is an extension of Gessel identity, quoted in Appendix A, to the case
of supersymmetric Schur polynomials. Notice that our unitary matrix model, (%
\ref{Z2}) above, is of this type since a principal specialization of the $x$
and $y$ set of variables $x_{i}=y_{i}=q^{i-1/2}$ $(i=1,...,N)$ and the
semiclassical limit of a principal specialization of the $z$ and $w$
variables, namely $z_{j}=w_{j}=-e^{-m}$ $(j=1,...N_{f})$ gives for the
r.h.s. of (\ref{HS})%
\begin{equation*}
\int_{(0,2\pi ]^{N}}\frac{d^{N}\mu }{(2\pi )^{N}}\frac{\prod_{i<j}4\sin ^{2}(%
\frac{1}{2}(\mu _{i}-\mu _{j}))\ \prod_{j}\theta _{3}^{\left( N\right)
}(e^{i\mu _{j}},q)}{\prod_{i}\left( 4\cos (\frac{1}{2}(\mu _{i}+im))\cos (%
\frac{1}{2}(\mu _{i}-im))\right) ^{N_{f}}},
\end{equation*}%
where $\theta _{3}^{(N)}(e^{i\mu },q)$ denotes a truncated theta function 
\cite{Foata}%
\begin{equation*}
\theta _{3}^{(N)}(z,q)=\,\sum_{n=-N}^{N}\ \left[ 
\begin{matrix}
2N \\ 
n+N%
\end{matrix}%
\right] _{q}\,q^{n^{2}/2}\,z^{n}=\,\left( \sqrt{q}\,z\,;q\right)
_{N}\,\left( \sqrt{q}\,z^{-1}\,;\,q\right) _{N}.
\end{equation*}%
If we consider the sum over all representations $\lambda $ in (\ref{HS})
(i.e. without the restriction $\lambda _{1}\leq N)$ as in a 2d Yang-Mills
theory and take $k_{1}\rightarrow \infty $ and $k_{2}\rightarrow \infty $
(while keeping $l_{1}=l_{2}=N_{f}$) then we have%
\begin{eqnarray}
\sum_{\lambda }\,e^{-2m\left\vert \lambda \right\vert }\mathrm{sdim}%
_{q}^{2}\lambda &=&\,\int_{(0,2\pi ]^{\infty }}\prod\limits_{k=1}^{\infty }%
\frac{\ \prod\nolimits_{j=1}^{\infty }\left( 1+q^{j-\frac{1}{2}}\mathrm{e}%
^{i\mu _{k}}\right) \left( 1+q^{j-\frac{1}{2}}\mathrm{e}^{-i\mu _{k}}\right) 
}{\left( 1+e^{-m}\mathrm{e}^{i\mu _{k}}\right) ^{N_{f}}\left( 1+e^{-m}%
\mathrm{e}^{-i\mu _{k}}\right) ^{N_{f}}}\frac{d\mu _{k}}{2\pi }%
\prod_{i<j}4\sin ^{2}(\frac{1}{2}(\mu _{i}-\mu _{j}))  \notag \\
&=&\lim_{N\rightarrow \infty }\widehat{Z}_{N_{f}}^{U(N)}=\frac{1}{\left(
1-e^{-2\left\vert m\right\vert }\right) ^{N_{f}^{2}}}%
\prod_{j=1}^{N}(1-q^{j})^{-j}(1-q^{j-\frac{1}{2}}e^{-\left\vert m\right\vert
})^{2N_{f}},  \label{rel}
\end{eqnarray}%
where $\widehat{Z}_{N_{f}}^{U(N)}\equiv \left( 2\pi /g\right)
^{N/2}e^{NN_{f}\left\vert m\right\vert }\widetilde{Z}_{N_{f}}^{U(N)}/\left(
(q;q)_{\infty }\right) ^{N}$ and we have defined the \textquotedblleft
supersymmetric half-$q$-deformed" dimensions \footnote{%
These dimensions can be seen as a $q$-deformation of the $t$-dimensions in 
\cite{MV}. In addition, the ring of $q$-superdimensions and its appearance
in $U\left( m\right\vert n)$ Chern-Simons theory were studied in \cite%
{Bourdeau:1991hn}. It remains to be analyzed whether the definition in \cite%
{Bourdeau:1991hn} is also given by a specialization of the supersymmetric
Schur polynomial.}%
\begin{equation*}
\mathrm{sdim}_{q}\lambda \equiv \,\mathrm{HS}_{\lambda
}(q^{1/2},q^{3/2},...|-1,...,-1).
\end{equation*}%
Notice that in (\ref{rel}), as usual with this type of description, we do
not have the full CS partition function part and a $\left( (q;q)_{\infty
}\right) ^{N}$ piece has to be added. Likewise, the factor $%
e^{-NN_{f}\left\vert m\right\vert }$ in (\ref{massive}) also does not appear
in the r.h.s. of (\ref{rel}) because such a term comes from the numerical
pre-factor in (\ref{symbol}) and the symbol that arises from writing the
l.h.s. of (\ref{rel}) as a Toeplitz determinant only gives the $z$-dependent
part of (\ref{symbol}) without numerical pre-factors.

As we shall see below, the massless case can also be analyzed with an
extension of Szeg\"{o}'s theorem. We collect here, for comparison with (\ref%
{rel}), the ensuing result%
\begin{eqnarray*}
\sum_{\lambda }\,\mathrm{sdim}_{q}^{2}\lambda &=&\int_{(0,2\pi ]^{\infty
}}\prod\limits_{k=1}^{\infty }\frac{\ \prod\nolimits_{j=1}^{\infty }\left(
1+q^{j-\frac{1}{2}}\mathrm{e}^{i\mu _{k}}\right) \left( 1+q^{j-\frac{1}{2}}%
\mathrm{e}^{-i\mu _{k}}\right) }{\left( 1+e^{-m}\mathrm{e}^{i\mu
_{k}}\right) ^{N_{f}}\left( 1+e^{-m}\mathrm{e}^{-i\mu _{k}}\right) ^{N_{f}}}%
\frac{d\mu _{k}}{2\pi }\prod_{i<j}4\sin ^{2}(\frac{1}{2}(\mu _{i}-\mu _{j}))
\\
&=&\lim_{N\rightarrow \infty }\widehat{Z}_{N_{f}}^{U(N)}(m=0)=\frac{%
G^{2}(1+N_{f})}{G(1+2N_{f})}N^{N_{f}^{2}}%
\prod_{j=1}^{N}(1-q^{j})^{-j}(1-q^{j-\frac{1}{2}})^{2N_{f}},
\end{eqnarray*}%
where $G$ is the Barnes $G$-function \cite{Barnes}, whose main definition
and properties are collected in the Appendix.


\subsection{Dual symbol}


Let us prove that there is also a dual symbol. This gives an alternative,
equivalent matrix model description, in analogy to the case of pure
Chern-Simons theory on $\mathbb{S}^{3}$ \cite{Szabo:2011eg,Szabo:2013vva,my}%
. In addition, it also justifies the use of the Fisher-Hartwig formalism
below, for the massless case. We have mentioned above the duality between
the symbols (\ref{S1}) and (\ref{S2}). This result is ultimately due to the
existence and equivalence of the Jacobi-Trudi formula and its dual (also
known as Nagelsbach-Kostka formula) \cite{Mac}%
\begin{equation*}
s_{\lambda }\left( x_{1},...,x_{N}\right) =\det \left( h_{\lambda
_{i}+j-i}\right) _{i,j=1}^{N}=\det \left( e_{\lambda _{i}^{^{\prime
}}+j-i}\right) _{i,j=1}^{N},
\end{equation*}%
where $h_{\lambda }$ and $e_{\lambda }$ are homogeneous and elementary
symmetric polynomials \cite{Mac}, respectively. These determinantal
expressions can also be interpreted as another definition of Schur
polynomials, alternative to the one given in Appendix A. The same result
holds for the supersymmetric Schur polynomials, replacing the homogeneous
and elementary symmetric functions with its supersymmetric counterparts \cite%
{MV}%
\begin{equation}
\mathrm{HS}_{\lambda }(x|z)=\det \left( h_{\lambda _{i}+j-i}(x|z)\right)
_{i,j=1}^{N}=\det \left( e_{\lambda _{i}^{^{\prime }}+j-i}(x|z)\right)
_{i,j=1}^{N},  \label{sgen1}
\end{equation}%
where the generating function of the supersymmetric homogenous and
elementary symmetric functions is now \cite{MV}%
\begin{equation}
\sum\limits_{r\geq 0}h_{r}(x|z)t^{r}=\frac{\prod_{j=1}^{l_{1}}\left(
1+z_{j}t\right) }{\prod_{i=1}^{k_{1}}\left( 1-x_{j}t\right) }\text{ and }%
\sum\limits_{r\geq 0}e_{r}(x|z)t^{r}=\frac{\prod_{i=1}^{k_{1}}\left(
1+x_{j}t\right) }{\prod_{j=1}^{l_{1}}\left( 1-z_{j}t\right) }.  \label{sgen2}
\end{equation}%
Hence, it immediately holds that the symbol and its dual are%
\begin{eqnarray*}
\varphi \left( z\right) &=&\frac{\prod_{j=1}^{k_{1}}\left( 1+x_{j}z\right)
\prod_{j=1}^{k_{2}}\,\left( 1+y_{j}/z\right) \,{\,}}{\prod_{j=1}^{l_{1}}%
\left( 1-z_{j}z\right) \,\prod_{j=1}^{l_{2}}\left( 1-w_{j}/z\right) }, \\
\widetilde{\varphi }\left( z\right) &=&\frac{\prod_{j=1}^{l_{1}}\left(
1+z_{j}z\right) \,\prod_{j=1}^{l_{2}}\left( 1+w_{j}/z\right) }{%
\prod_{j=1}^{k_{1}}\left( 1-x_{j}z\right) \prod_{j=1}^{k_{2}}\,\left(
1-y_{j}/z\right) }.
\end{eqnarray*}%
After the principal specialization $x_{j}=y_{j}=q^{j-1/2}$ and $%
z_{j}=w_{j}=e^{-m}$ with $l_{1}=l_{2}=N_{f}$ and $k_{1}=k_{2}\rightarrow
\infty $ we have that%
\begin{equation}
\varphi \left( z\right) =\frac{\theta _{3}({z},q)}{\left( 1+e^{-m}/z\right)
^{N_{f}}\left( 1+e^{-m}z\right) ^{N_{f}}}\text{ \ and \ }\widetilde{\varphi }%
\left( z\right) =\frac{\left( 1-e^{-m}/z\right) ^{N_{f}}\left(
1-e^{-m}z\right) ^{N_{f}}}{\theta _{3}(-{z},q)},  \label{symbeq}
\end{equation}%
where $\varphi \left( z\right) =e^{mN_{f}}\varphi _{\mathrm{CSM}}(z)$
(recall (\ref{symbol})). The numerical factor $e^{mN_{f}}$ does not appear
in any of the symbols in (\ref{symbeq}) because it comes out of the
relationship (\ref{trig}). Notice also how consideration of Szeg\"{o}'s
theorem confirms that the determinant for both cases in (\ref{symbeq})
coincides. It is also worth mentioning that, while the two symbols give the
same partition function, if one studies Wilson loops in a representation $%
\lambda $, then $\left\langle W_{\lambda }\right\rangle _{\varphi \left(
z\right) }=\left\langle W_{\lambda ^{^{\prime }}}\right\rangle _{\widetilde{%
\varphi }\left( z\right) }$. This is shown explicitly for the pure
Chern-Simons case in \cite{my} and the same proof again follows here with
the use of (\ref{sgen1}) and (\ref{sgen2}) instead of their
non-supersymmetric versions. Alternatively, notice that it holds that $%
\mathrm{HS}_{\lambda }(x|z)=\mathrm{HS}_{\lambda ^{\prime }}(z|x)$.


\section{Massless case}


While the massive case can be analyzed with the strong Szeg\"{o} theorem and
with generalized Cauchy identities, the massless case develops a
Fisher-Hartwig singularity \cite{FH,rev}. This is the only particular case
of our problem where the Cauchy identity and Szeg\"{o}'s theorem is not
directly applicable since the situation where the symbol of the Toeplitz
determinant (weight function of the unitary matrix model) has a
zero/singularity on $\mathbb{S}^{1}$ is well-known to require Fisher-Hartwig
(FH)\ asymptotics \cite{FH,rev}, which refines the strong Szeg\"{o} theorem.


\subsection{$g=\infty $ limit case}


In this particular case, we do not have the Gaussian/theta function part and
we end up with a Toeplitz determinant whose symbol has just one FH
singularity. The Cauchy identity diverges in this case because it
corresponds to the specialization $x_{i}=y_{i}=1$ for $i=1,...N.$

This corresponds to the absence of a Chern-Simons term, a case which has
been studied in \cite{Safdi:2012re} but for large N and in the setting of a
more general matter content, where the matrix model is characterized by
double sine functions. In the Appendix A of \cite{Kapustin:2010mh}, the
massive case (with different masses) without Chern-Simons term is also
studied, and their resulting formula is nothing else but the Cauchy
determinant. As explained above, the massless case is outside the domain of
convergence of such formula.

Taking into account the duality between symbols discussed above we can
directly use the result in \cite{oneFH1,oneFH2}, which computes the matrix
model (\ref{HSid}) for finite $N$ for a symbol $\phi (z)=(1-z)^{\alpha
}(1-z^{-1})^{\beta }$, giving the result 
\begin{equation}
\det T_{N}\left( \phi \right) =G(N+1)\frac{G(\alpha +\beta +N+1)}{G(\alpha
+\beta +1)}\frac{G(\alpha +1)}{G(\alpha +N+1)}\frac{G(\beta +1)}{G(\beta
+N+1)},  \label{1}
\end{equation}%
where $G(z)$ is again Barnes G-function. Then, if $\alpha =\beta =N_{f}$,
then $\det T_{N}\left( \phi \right) =\widehat{Z}_{N_{f}}^{U(N)}$ and we have 
\begin{equation}
\widehat{Z}_{N_{f}}^{U(N)}(m=0,g=\infty )=G(N+1)\frac{G(2N_{f}+N+1)}{%
G(2N_{f}+1)}\frac{G^{2}(N_{f}+1)}{G^{2}(N_{f}+N+1)}.  \label{massless}
\end{equation}%
We note that consideration of Selberg integral also leads to (\ref{massless}%
) \cite{rev}. Notice that this is in principle very different from the
massive case, which is given by Cauchy identity, even for $N$ finite:%
\begin{equation*}
Z_{N}=\prod_{i,j=1}^{N_{f}}\frac{1}{1-x_{i}y_{j}}=\frac{1}{\left(
1-e^{-2\left\vert m\right\vert }\right) ^{N_{f}^{2}}}\text{ valid for }N\geq
N_{f}\text{.}
\end{equation*}%
The large $N$ limit of (\ref{massless}) is very well-known%
\begin{equation*}
\widehat{Z}_{N_{f}}^{U(N)}(m=0,g=\infty )=\frac{G^{2}(1+N_{f})}{G(1+2N_{f})}%
N^{N_{f}^{2}}\text{ for }N\rightarrow \infty .
\end{equation*}%
This will be a piece of the large $N$ result of the massless case with $g$
finite, as we shall see below. Note that this $g=\infty $ limit for the
massless case cannot be connected with the $g=\infty $ limit of section 3,
where we assumed that $m$ is also large and scales with $g$.


\subsection{Large N}


We can keep the Gaussian/theta function part and use the result on
Fisher-Hartwig (FH) asymptotics, which is a generalization of Szeg\"{o}'s
result \cite{FH}. Note that above we used instead an exact result for finite 
$N$. The symbols of FH class have the following form \cite{rev} 
\begin{equation}
f(z)=e^{V(z)}\,z^{\sum_{j=0}^{m}\,\beta
_{j}}\prod_{j=0}^{m}|z-z_{j}|^{2\alpha _{j}}\,g_{z_{j},\,\beta
_{j}}\,(z)\,z_{j}^{-\beta _{j}},\qquad z=e^{i\theta },\quad 0\leq \theta
<2\pi ,  \label{fFH}
\end{equation}%
for some $m=0,1,2,\dots ,$ with 
\begin{align}
& z_{j}=e^{i\,\theta _{j}},\qquad j=0,1,\dots ,m,\qquad 0=\theta _{0}<\theta
_{1}<\dots <\theta _{m}<2\pi ,  \label{eq75} \\
& g_{z_{j}\,\beta _{j}}(z)\equiv g_{\beta _{j}}(z)=e^{i\pi \beta _{j}}\text{
for }0\leq \arg z<\theta _{j}\text{ (}e^{-i\pi \beta _{j}}\text{ otherwise).}
\label{eq76} \\
& \Re \left( \alpha _{j}\right) >-\frac{1}{2},\qquad \beta _{j}\in 
\mathbb{C}
,\qquad j=0,1,\dots ,m,  \label{eq77}
\end{align}%
and $V\!(e^{i\theta })$ is a sufficiently smooth function on $\mathbb{S}^{1}$%
. Here the condition on $\Re \left( \alpha _{j}\right) $ guarantees
integrability. Note that a FH singularity at $z_{j},$ $j=1,\dots ,m$,
consists of a root-type singularity%
\begin{equation}
|z-z_{j}|^{2\alpha _{j}}=|2\sin \frac{\theta -\theta _{j}}{2}|^{2\alpha _{j}}
\label{eq78}
\end{equation}%
and a jump singularity $z^{\beta _{j}}\,g_{\beta _{j}}(z)$ at $z_{j}$ (note
that $z^{\beta _{j}}\,g_{\beta _{j}}(z)$ is continuous at $z=1$ for $j\neq 0$%
). Notice that the symbol $\widetilde{\varphi }\left( z\right) $ in (\ref%
{symbeq}) is of this type with $m=0$ and hence with only one FH singularity
of the root-type, because $\alpha _{0}=N_{f}$ and $\beta _{0}=0$. The
asymptotic form of $\det T_{N}\left( f\right) $ for the general symbol above
is,\label{eq84} 
\begin{eqnarray}
\det T_{N}\left( f\right) &=&E(e^{V},\,\alpha _{0},\dots ,\alpha
_{m},\,\beta _{0},\dots ,\beta _{m},\,\theta _{0},\dots ,\theta
_{m})\,n^{\sum_{j=0}^{m}(\alpha _{j}^{2}-\beta
_{j}^{2})}\,e^{NV_{0}}(1+o(1)),\qquad  \notag \\
V_{0} &=&\frac{1}{2\pi }\int_{0}^{2\pi }V\!(e^{i\theta })d\theta
\end{eqnarray}%
as $N\rightarrow \infty $. For a Fisher-Hartwig symbol, in addition we have 
\begin{equation}
E(e^{V},\,\alpha _{0},\dots ,\alpha _{m},\,\,\theta _{0},\dots ,\theta
_{m})=E(e^{V})\prod_{0\leq j<k\leq m}|e^{i\theta _{j}}-e^{i\theta
_{k}}|^{-2\alpha _{j}\alpha _{k}}\prod_{j=0}^{m}e^{-\alpha _{j}\widehat{V}%
(e^{i\theta _{j}})}\times \prod_{j=0}^{m}E_{\alpha _{j}}  \label{E1}
\end{equation}%
where%
\begin{alignat}{3}
& E(e^{V})=e^{\sum_{k=1}^{\infty }k\,V_{k}\,V_{-k}},\qquad & V_{k}& =\text{
Fourier coefficient of }V\!(e^{i\theta }), & &  \label{eq86} \\
& \widehat{V}\!(e^{i\theta _{j}})=V\!(e^{i\theta _{j}})-V_{0} & & & &
\label{eq87} \\
& E_{\alpha _{j}}=G^{2}(1+\alpha _{j})/G(1+2\alpha _{j}),\qquad & & & &
\label{eq88}
\end{alignat}%
Notice that the term (\ref{eq86}) is the content of Szeg\"{o}'s theorem, the
rest therefore extends it with additional contributions. Note also that (\ref%
{eq88}) is essentially the large $N$ limit of the finite $N$ result above (%
\ref{1}). Since we only have one FH singularity ($j=0$), the product term in
(\ref{E1}) does not contribute. Taking into account that our symbol is (\ref%
{symbeq}) with $m=0$ then $\alpha _{0}=N_{f}$, $\beta _{0}=0$ and $z_{0}=1$ (%
$\theta _{0}=0$). Therefore, we obtain%
\begin{equation}
\widetilde{Z}_{N_{f}}^{U(N)}(m=0)=\left( \frac{g}{2\pi }\right) ^{N/2}\frac{%
G^{2}(1+N_{f})}{G(1+2N_{f})}N^{N_{f}^{2}}\prod\limits_{j=1}^{\infty }\left(
1-q^{j}\right) ^{N-j}\left( 1-q^{j-\frac{1}{2}}\right) ^{2N_{f}}\text{ for }%
N\rightarrow \infty .  \label{largeN}
\end{equation}


\subsection*{Acknowledgements}


We would like to thank M. Mari\~{n}o, M. Yamazaki and K. Zarembo for useful
comments. JR acknowledges financial support from projects FPA 2010-20807.
GAS acknowledges financial support from CONICET PIP2010-0396 and
PIP2013-0595. GAS would like to thank the High energy group of ICTP and
Department ECM of Universitat de Barcelona for hospitality during the early
stages of this work. MT acknowledges financial support from a Juan de la
Cierva Fellowship, from MINECO (grant MTM2011-26912), from the European
CHIST-ERA project CQC (funded partially by MINECO grant
PRI-PIMCHI-2011-1071) and from Comunidad de Madrid (grant QUITEMAD+-CM, ref.
S2013/ICE-2801). The department ECM of Universitat de Barcelona is also
thanked for warm hospitality.

\appendix

\section{Mathematical identities}

\label{App:AppendixA}

We collect here a number of mathematical identities and results used through
the text. A partition is a finite sequence of nonnegative integers $\lambda
_{1}\geq \cdot \cdot \cdot \geq \lambda _{n}\geq 0$. Associated to every
partition is a Young diagram with $\lambda _{i}$ squares in the $i$-th row
and the rows are understood to be aligned on the left. There is a unique $n$
such that $\lambda _{n}>0$ but $\lambda _{n+1}=0$ and this $n=l(\lambda )$
is the length of $\lambda $. The number $\left\vert \lambda \right\vert
=\sum\nolimits_{i}$ $\lambda _{i}$ is called the size of $\lambda$ and we
denote by $\lambda^{\prime }$ the conjugate partition to $\lambda$. Schur
polynomials $s_\lambda(x)$ are $|\lambda|$-th homogeneous symmetric
polynomials, if $\lambda $ is any partition of length $n$ we define \cite%
{Mac}%
\begin{equation*}
s_{\lambda }(x_{1},...,x_{n}):=\frac{\det \left( x_{j}^{\lambda
_{k}+n-k}\right) _{j,k=1}^{n}}{\det \left( x_{j}^{n-k}\right) _{j,k=1}^{n}}.
\end{equation*}%
We shall summarize now a number relationships between Schur polynomials and
Toeplitz determinants (equivalently, unitary matrix models \cite%
{Szabo:2010sd}-\cite{Szabo:2013vva}). We begin with two classical results by
Gessel and Baxter which are relevant in Section 4.

\subsection{Gessel and Baxter identities}

We first quote Gessel's formula for the product of Schur polynomials in
terms of a Toeplitz determinant, which reads~\cite{Gessel}%
\begin{equation}
\sum_{\lambda \,;l(\lambda )\leq N}\,s_{\lambda }\left( x\right)
\,s_{\lambda }\left( y\right) =\det (A_{i-j})_{i,j=1}^{N}\ ,
\label{Toeplitz}
\end{equation}%
where 
\begin{equation}
A_{i}=A_{i}(x,y)=\sum_{l=0}^{\infty }\,\mathfrak{h}_{l+i}(x)\,\mathfrak{h}%
_{l}(y)\ ,
\end{equation}%
are the Fourier coefficients of the symbol (entries of the Toeplitz matrix)
and $\mathfrak{h}_{r}(x)$ is the $r$-th homogeneous symmetric function,
characterized by its generating function $\sum\nolimits_{r\geq 0}\mathfrak{h}%
_{r}t^{r}=\prod\nolimits_{j\geq 1}\left( 1-x_{j}\,t\right) ^{-1}$. The
symbol of the Toeplitz determinant is then \cite{Gessel,TW}%
\begin{equation}
\varphi \left( z\right) =\sum_{i=-\infty }^{\infty
}\,A_{i}(x,y)\,z^{i}=\prod\limits_{j\geq 1}\,\left( 1-y_{j}\,z^{-1}\right)
^{-1}\,\left( 1-x_{j}\,z\right) ^{-1}\ .
\end{equation}%
The dual version is \cite{TW}%
\begin{equation}
\sum_{\lambda \,:\,\lambda _{1}\leq N}\,s_{\lambda }\left( x\right)
\,s_{\lambda }\left( y\right) =\det (\widetilde{A}_{i-j})_{i,j=1}^{N},
\label{Toep2}
\end{equation}%
where the Fourier coefficients are now in terms of elementary symmetric
functions and the symbol is%
\begin{equation}
\widetilde{\varphi }\left( z\right) =\sum_{i=-\infty }^{\infty }\,\widetilde{%
A}_{i}(x,y)\,z^{i}=\prod\limits_{j\geq 1}\,\left( 1+y_{j}\,z^{-1}\right)
\,\left( 1+x_{j}\,z\right) \ .
\end{equation}%
Notice that the restriction in the sum over representations in (\ref%
{Toeplitz}) is a bound on the size of the first column whereas in (\ref%
{Toep2}) the first row is bounded by $N$. When the sum is not restricted,
the Toeplitz determinants (equivalently, the unitary matrix models) are in
principle infinite-dimensional and the Cauchy-Binet identity holds:%
\begin{equation*}
\sum_{\lambda }\,s_{\lambda }\left( x_{1},...,x_{p}\right) \,s_{\lambda
}\left( y_{1},...,y_{q}\right) =\prod_{i=1}^{p}\prod_{j=1}^{q}\frac{1}{%
1-x_{i}y_{j}},
\end{equation*}%
where the products goes from $1$ to the number of $x$ and $y$ variables.
Schur polynomials satisfy the property $s_{\lambda }(x_{1},...,x_{n})=0$ if $%
l(\lambda )>n$ \cite{Mac}, which implies a truncation of the sum for a
finite number of variables of the Schur polynomials. Thus, the sum on the
l.h.s. is effectively over all partitions $\lambda $ of length$\leq \min
(p,q)$. This result is translated into an statement for Toeplitz
determinants by the following Lemma:

\begin{lemma}
(Baxter \cite{Bax}, Lemma 7.4) Let $D_{n}(\sigma )$ denote the determinant
of a Toeplitz matrix $n\times n$ and symbol $\sigma $, then 
\begin{equation*}
D_{n}\left( \sigma \right) =\Pi _{i,j}\left( 1-\alpha _{i}\beta _{j}\right)
^{-1},
\end{equation*}%
where the symbol is, specifically $\sigma \left( z\right) =\Pi
_{i=1}^{k}\left( 1-\alpha _{i}z\right) \Pi _{i=1}^{m}\left( 1-\beta
_{j}z^{-1}\right) $. This result is valid for $n\geq \max (k,m)$ and
independent of n.
\end{lemma}

Notice again that Cauchy-Binet identity only says that $\lim_{n\rightarrow
\infty }D_{n-1}\left( \sigma \right) =\Pi _{i,j}\left( 1-\alpha _{i}\beta
_{j}\right) ^{-1}$, but for this symbol, the determinant will give the same
result for any finite size, from infinite size, down to the number of
product terms in the symbol. This result leads to the last identity in (\ref%
{SC}).

\subsection{Cauchy -Binet identity and Szeg\"{o}'s theorem}

Notice that, at least in our context, the statement of Szeg\"{o}'s theorem
is equivalent to the Cauchy-Binet formula (\ref{SC}) when the latter is
written in Miwa variables \cite{vM}%
\begin{equation}
\sum_{\lambda }\,s_{\lambda }(x)\,s_{\lambda }(y)=\exp \,\left( \sum_{k\geq
1}\,k\,m_{k}\,t_{k}\right) \,,  \label{Sc}
\end{equation}%
where%
\begin{equation*}
m_{k}=\frac{1}{k}\,\sum_{i\geq 1}\,x_{i}^{k}\qquad \mbox{and}\qquad t_{k}=%
\frac{1}{k}\,\sum_{i\geq 1}\,y_{i}^{k}
\end{equation*}%
are power sums of the sets of variables $x$ and $y$. Hence, the construction
of Miwa variables is equivalent to the computation of the moments of the
logarithm of the symbol, which is the potential of the matrix model (coming
from the Taylor expansion of a logarithm).

\subsection{Barnes G-function}

The Barnes G-function \cite{Barnes} is a double-Gamma function that can be
for example defined with the functional equation 
\begin{equation*}
G(z+1)=\Gamma \left( z\right) G(z)
\end{equation*}%
with normalization $G(1)=1$. Its asymptotic expansion is especially useful%
\begin{equation}
\ln G(t+a+1)=\frac{1}{12}-\ln A-\frac{3t^{2}}{4}-at+\frac{t+a}{2}\ln (2\pi
)+(\frac{t^{2}}{2}+at+\frac{a^{2}}{2}-\frac{1}{12})\ln t+o(t^{-1}),\text{ \
as }\;t\rightarrow \infty .  \notag
\end{equation}


\section{Moment problem and discretization of the matrix model}

\label{App:AppendixB}

In this paper, we have studied a one matrix model with potential%
\begin{equation}
V(z)=\frac{1}{2g}\ln ^{2}z+N_{f}\ln \left( 1+z_{i}\frac{e^{m}}{c}\right)
\left( 1+z_{i}\frac{e^{-m}}{c}\right) ,  \label{V}
\end{equation}%
where $z\in \left( 0,\infty \right) $. Thus, the confining properties are
those of the Stieltjes-Wigert potential \cite{Tierz} since, for large $z,$
the first term in (\ref{V}) dominates. Therefore, we expect the model to be
associated to an undetermined moment problem, as happens with the
Stieltjes-Wigert matrix model \cite{Tierz,deHaro:2005rz,Szabo:2013vva}. This
means that there are infinitely many deformations of the measure (\ref%
{measure}) with identical orthogonal polynomials $p_{n}(z)$ and therefore
identical (\ref{pols}). In consequence, every matrix model constructed from
such a measure possesses the same partition function.

This is demonstrated by considering Krein's proposition \cite{Tierz}, which
gives a sufficient condition for a moment problem to be undetermined. The
condition is for the weight function $\omega (z)=\exp (-V(z))$ to satisfy%
\begin{equation*}
-\int_{0}^{\infty }\frac{\ln \omega (z)}{(1+z)}\frac{dz}{\sqrt{z}}<\infty .
\end{equation*}%
The integral converges for our potential (\ref{V}), as it happens in the
pure Stieltjes-Wigert case \cite{Tierz}, and hence the moment problem
associated is undetermined. Alternatively, this can be seen even more
explicitly by following Stieltjes directly \cite{Tierz}, by showing%
\begin{equation*}
\int_{0}^{\infty }z^{k}e^{-\frac{1}{2g}\ln ^{2}z+N_{f}\ln \left( 1+z_{i}%
\frac{e^{m}}{c}\right) \left( 1+z_{i}\frac{e^{-m}}{c}\right) }\sin \left(
2\pi \ln z/\ln q\right) =0,
\end{equation*}%
which follows, as happens in the case $N_{f}=0$, by the change of variables $%
v=-(k+1)/2+\ln z$, the periodicity of $\sin ($\textperiodcentered $)$, and
the fact that $\sin $ is an odd function. Thus for any $\theta \in \left[
-1,1\right] $ the weight function $\omega_\theta (z)=\omega (z)(1+\theta
\sin \left( 2\pi \ln z/\ln q\right) )$, where $\omega (z)=\exp (-V(z))$ and $%
V(z)$ is (\ref{V}), has the same positive integer moments as $\omega (z)$
and therefore the corresponding (infinitely many) matrix models have the
same partition function.

The set of all solutions to an indeterminate moment problem always includes
discrete measures (the so-called canonical solutions of a moment problem are
discrete measures), which implies that there is a discrete matrix model
equivalent to the continuous one. In the case of the Stieltjes-Wigert matrix
model, the discrete matrix model is known explicitly since the discrete
measure with the same moments as $e^{-\frac{1}{2g}\ln ^{2}z}$ is known to be 
$M\left( q\right) \sum\nolimits_{n\in 
\mathbb{Z}
}q^{n^{2}/2+n}\delta \left( x-q^{n}\right) $ with $M(q)$ a suitable constant
(see \cite{deHaro:2005rz,Szabo:2013vva} and references therein). The
analogous result for (\ref{V}) is not known and not immediate to obtain.
Hence to find the explicit form of the discrete matrix model which is
equivalent to (\ref{CSM}) and, after the change of variables, to (\ref{Z}),
is an open problem.

Notice however that a straightforward discretization of the Mordell integral
gives already very good results for large coupling constant $g$ since it is
known that, for the integral \cite{CS}%
\begin{equation*}
\varphi (g,c,z)=\int_{-\infty }^{\infty }\frac{e^{-\frac{1}{2g}(x-z)^{2}}}{%
e^{cx}+1}dx,
\end{equation*}%
the straightforward discretization which is standard trapezoidal quadrature
with step $h$, that is%
\begin{equation*}
\varphi (g,c,z)=h\sum_{k=-\infty }^{\infty }\frac{e^{-\frac{1}{2g}(kh-z)^{2}}%
}{e^{ckh}+1}+\mathcal{E}\left( h\right) ,
\end{equation*}%
has an error term which is bounded by%
\begin{eqnarray*}
\left\vert \mathcal{E}\left( h\right) \right\vert &\leq &{2\exp (\pi
^{2}/8c^{2}g-\pi ^{2}/hc)\text{ for }g>\frac{h}{4c},} \\
\left\vert \mathcal{E}\left( h\right) \right\vert &\leq &{2\exp (-2\pi
^{2}g/h^{2})\text{ for\ }g\leq \frac{h}{4c}}\text{ }.
\end{eqnarray*}%
The contour integral result in \cite{CS} allows for the generalization to
the case corresponding to $N_{f}>1$.


\end{document}